\documentclass[]{raa}           
\usepackage{graphicx,times}
\usepackage{natbib}
\usepackage{amssymb,amsmath}
\bibpunct{(}{)}{;}{a}{}{,}

\usepackage[a4paper=true,pagebackref=true]{hyperref}
\usepackage{longtable}
\usepackage{footnote}
\usepackage{threeparttable}
\usepackage{tabularx}
\usepackage{adjustbox}
\usepackage{multirow}

\hypersetup{pdftitle = The title of my PDF, pdfauthor = My name, pdfsubject= The subject, pdfkeywords = keyword1 keyword2 keyword3} 
\hypersetup{colorlinks = true, linkcolor = green, anchorcolor = red, citecolor = blue, filecolor = red, pagecolor = red, urlcolor = red}

\begin{document}

   \title{A Study of Multiple Molecular Lines at the 3 mm Band toward Gas Infalling Sources
}

   \volnopage{Vol.0 (20xx) No.0, 000--000}      
   \setcounter{page}{1}          

   \author{Yang Yang
      \inst{1,2,3}
   \and Xi Chen
      \inst{2}
   \and Zhibo Jiang 
      \inst{3,4}
   \and Zhiwei Chen
      \inst{3}
   \and Shuling Yu
      \inst{3,5}
   \and Jun Li
      \inst{2}
   }

   \institute{College of Elementary Education, Changsha Normal University, Changsha 410100, China; {\it yangy@csnu.edu.cn}\\
        \and
             Center for Astrophysics, Guangzhou University, Guangzhou 510006, China
        \and 
             Purple Mountain Observatory, Chinese Academy of Sciences, Nanjing 210023, China
        \and
             China Three Gorges University, Yichang, Hubei 443002, People's Republic of China\\
        \and
             University of Science and Technology of China, Chinese Academy of Sciences, Hefei, Anhui 230026, People's Republic of China\\
\vs\no
   {\small Received~~20xx month day; accepted~~20xx~~month day}}

\abstract{The study of multiple molecular spectral lines in gas infalling sources can provide the physical and chemical properties of these sources and help us estimate their evolutionary stages. We report line detections within the 3 mm band using the FTS wide-sideband mode of the IRAM 30 m telescope toward 20 gas-infalling sources. Using XCLASS, we identify the emission lines of up to 22 molecular species (including a few isotopologues) and one hydrogen radio recombination line in these sources. H$^{13}$CO$^+$, HCO$^+$, HCN, HNC, c-C$_3$H$_2$, and CCH lines are detected in 15 sources. We estimate the rotation temperatures and column densities of these molecular species using the LTE radiative transfer model, and compare the molecular abundances of these sources with those from nine high-mass star-forming regions reported in previous studies and with those from the chemical model. Our results suggest that G012.79-0.20, G012.87-0.22 clump A and B, and G012.96-0.23 clump A may be in the high-mass protostellar object stage, while sources with fewer detected species may be in the earlier evolutionary stage. Additionally, the CCH and c-C$_3$H$_2$ column densities in our sources reveal a linear correlation, with a ratio of N(CCH)/N(c-C$_3$H$_2$) = 89.2$\pm$5.6, which is higher than the ratios reported in the literature. When considering only sources with lower column densities, this ratio decreases to 29.0$\pm$6.1, consistent with those of diffuse clouds. Furthermore, a comparison between the N(CCH)/N(c-C$_3$H$_2$) ratio and the sources' physical parameters reveals a correlation, with sources exhibiting higher ratios tending to have higher kinetic temperatures and H$_2$ column densities.
\keywords{Star formation --- ISM: clouds --- ISM: lines and bands --- ISM: molecules}
}

   \authorrunning{Y. Yang et al.}            
   \titlerunning{A Study of Molecular Lines at the 3 mm Band}  
   \maketitle

%
\section{Introduction}           
\label{sec:intro}

In the early stages of star formation, the gas within molecular clouds collapses inward under the gravitational action \citep{Bachiller+1996, Shu+etal+1987}. This gas infall motion accumulates material, contributing to the mass growth of the central objects. However, the mass growth of high-mass young stellar objects (YSOs) cannot be fully explained by simple gravitational collapse and accretion alone, as factors such as radiation pressure, turbulence, and magnetic fields also affect the mass growth of YSOs during the star-forming process \citep[e.g.][]{Bhadari+etal+2023}. Consequently, additional physical mechanisms are necessary to account for the mass increase of YSOs. Therefore, studying molecular clumps undergoing gas infall will enhance our understanding of the physical mechanisms involved in high-mass star formation.

Numerous studies have shown that the gas-phase molecular composition of clumps undergoing star formation is quite rich, especially in high-mass star-forming regions \citep[e.g.][]{Liu+etal+2020}. Among various molecular species, HCO$^+$, HNC, HCN, CS, and H$_2$CO can trace dense regions close to the core and serve as optically thick spectral lines for identifying infall sources \citep[e.g.][]{Klaassen+etal+2012, Contreras+etal+2018, Saral+etal+2018, Su+etal+2019}. The spectral line profiles of molecules, such as CO and HCO$^+$, which exceed the Gaussian profile wings, can also be used to trace gas outflows \citep[e.g.,][]{Mottram+etal+2017, Yang+etal+2018, Nagy+etal+2020, Yang+etal+2022}. SiO is considered to primarily form in shock environments, where silicon is liberated from dust grains and combines with oxygen to form SiO, making it an effective tracer of ongoing outflow activity \citep[e.g.,][]{Cunningham+etal+2018, Guerra-Varas+etal+2023}. Additionally, some carbon chain molecules are the best probes of the less chemically evolved cores, or chemically fresh material \citep[e.g.][]{Pineda+etal+2020}. The evolutionary time-scale of star-forming sources can be estimated by analyzing carbon-chain species such as C$_3$, C$_3$H$_2$, CH$_3$CCH, and CCH in molecular clumps and comparing their abundances with simulated values from chemical models \citep[e.g.,][]{Mookerjea+etal+2012}.

Our study focuses on gas infalling sources identified in previous studies. The redshifted self-absorption profiles of optically thick lines, commonly named as the blue profile \citep{Mardones+etal+1997}, are often used as an indirect tracer of gas infall. Concurrently, optically thin line data are used to trace the central radial velocity of these sources, helping to confirm that the blue profiles of the optically thick lines indicate gas infall rather than other effects, such as multiple velocity components. The sources selected for this study were detected through a blind search of all available $^{12}$CO (1-0) and its isotopologues $^{13}$CO (1-0) and C$^{18}$O (1-0) data from the Milky Way Imaging Scroll Painting (MWISP) project \citep[][hereafter referred to as Paper I]{Jiang+etal+2023}. We used the combinations of $^{12}$CO/$^{13}$CO and $^{13}$CO/C$^{18}$O as two pairs of optically thick and thin lines to trace the gas infall motions. Subsequently, we conducted further single-point and mapping observations of these infall candidates: The single-point observations were carried out using the Purple Mountain Observatory's 13.7 m telescope. We used more reliable infall tracers, HCO$^+$ (1-0) and HCN (1-0), to observe 133 infall candidates, further confirming the presence of infall signatures \citep[][hereafter referred to as Paper II]{Yang+etal+2020}. In \citet{Yang+etal+2021,Yang+etal+2023b}, referred to as Papers III and IV respectively, we used the IRAM 30 m telescope to conduct mapping observations of a combination of the optically thick line HCO$^+$ (1-0) and the optically thin line H$^{13}$CO$^+$ (1-0) for 37 infall candidates confirmed in Paper II, and identified gas infall motion in 29 of them. In these mapping observations, the Fourier Transform Spectrometer (FTS) in its wide-sideband mode covered multiple molecular line data in the frequency ranges of 83.7 -- 91.5 GHz and 99.4 -- 107.2 GHz. These data can be used to identify multiple molecular lines at the 3 mm band, as well as the spatial distribution of these molecular species.

In \citet{Yang+etal+2023a} (referred to as Paper V), we analyzed multiple spectral line data for nine sources exhibiting typical infall profiles. Among these sources, a total of 7 to 27 molecules and isotopologues transition lines were identified at the 3 mm band. Comparing the observed molecular abundances with the abundances obtained from chemical simulations revealed that most of these sources are in the early stage of high-mass protostellar object(HMPO). In this paper, we will proceed to present the mapping results for the remaining 20 infall sources, and analyze the spectral line data for these sources. A brief description of the mapping observations is given in Section \ref{sec:obs}, while in Section \ref{sec:result}, we present the identification of spectral lines and mapping results for these sources. In Section \ref{sec:analysis}, we identify clumps in these sources and analyze their physical properties, and use the eXtended CASA Line Analysis Software Suite (XCLASS) to conduct line fitting and estimate the rotation temperatures and column densities of the molecular species. We compare the molecular abundances with those of nine high-mass star-forming regions in our previous study. In addition, we give a comparison of CCH and c-C$_3$H$_2$ column densities. Finally, Section \ref{sec:summary} summarizes the results and analysis of this study.

\section{Sources and Observations}\label{sec:obs}

\begin{table*}
\begin{center}
  \caption{Source List.}\label{Tab:src}
 \setlength{\tabcolsep}{1mm}{
\begin{tabular}{ccccccccc}
  \hline\noalign{\smallskip}
Source  &  RA  &  Dec  & $v_{\text{LSR}}$ &  Distance  & RMS & Extent \\
Name &	(J2000)	&	(J2000)	&	(km s$^{-1}$)	&	(kpc) & (K)  & of map \\
  \hline\noalign{\smallskip}
G012.79-0.20 & 18:14:11.5 & -17:56:15 & 35.8 & 2.4$\pm$0.2 $^1$ &  0.18 & $4.5^{\prime}\times4.5^{\prime}$ \\
G012.87-0.22 & 18:14:26.4 & -17:52:43 & 35.4 & 2.4$\pm$0.2  $^1$ &  0.19 & $4.5^{\prime}\times4.5^{\prime}$ \\
G012.96-0.23 & 18:14:39.0 & -17:48:25 & 35.2 & 2.4$\pm$0.2  $^1$ &  0.19 & $2.5^{\prime}\times2.5^{\prime}$ \\
G014.25-0.17 & 18:17:02.2 & -16:38:40 & 38.1 & 3.3$_{-0.5}^{+0.4}$ $^2$ &  0.19 & $3^{\prime}\times3^{\prime}$ \\
G017.09+0.82 & 18:18:59.0 & -13:40:19 & 22.3 & 2.0$_{-0.6}^{+0.5}$ &  0.22 & $3^{\prime}\times3^{\prime}$ \\
G025.82-0.18 & 18:39:04.2 & -06:24:29 & 93.7 & 5.0$\pm$0.3  & 0.21 & $3^{\prime}\times3^{\prime}$ \\
G028.97+3.35 & 18:32:16.6 & -01:59:27 & 7.3 & 0.4$\pm$0.02 $^3$ &  0.13 & $3^{\prime}\times3^{\prime}$ \\
G029.06+4.58 & 18:28:04.2 & -01:20:47 & 7.5 & 0.5$_{-0.2}^{+0.5}$ &  0.16 & $2.5^{\prime}\times2.5^{\prime}$ \\
G030.17+3.68 & 18:33:18.1 & -00:46:31 & 9.0 & 0.6$\pm$0.5 &  0.13 & $2^{\prime}\times2^{\prime}$ \\
G031.41+5.25 & 18:29:58.6 & +01:02:31 & 8.3 & 0.6$\pm$0.5 &  0.12 & $2^{\prime}\times2^{\prime}$ \\
G036.02-1.36 & 19:01:57.5 & +02:07:55 & 31.7 & 2.0$\pm$0.4 &  0.17 & $4^{\prime}\times4^{\prime}$ \\
G037.05-0.03 & 18:59:06.3 & +03:38:41 & 81.4 & 4.9$\pm$0.5   & 0.19 & $4^{\prime}\times4^{\prime}$ \\
G049.07-0.33 & 19:22:42.3 & +14:10:00 & 60.7 & 4.7$\pm$0.8  $^2$ &  0.19 & $3^{\prime}\times3^{\prime}$ \\
G079.71+0.14 & 20:34:22.8 & +40:30:55 & 1.2 & 0.7$_{-0.7}^{+2.8}$ &  0.15 & $2.5^{\prime}\times2.5^{\prime}$ \\
G110.40+1.67 & 23:01:57.9 & +61:50:55 & -11.2 & 0.8$_{-0.7}^{+0.6}$ &  0.12 & $2.5^{\prime}\times2.5^{\prime}$ \\
G121.34+3.42 & 00:35:39.0 & +66:14:34 & -5.2 & 0.2$_{-0.2}^{+0.5}$ &  0.12 & $3^{\prime}\times3^{\prime}$ \\
G126.53-1.17 & 01:21:39.1 & +61:29:25 & -12.3 & 0.7$\pm$0.5 &  0.10 & $2^{\prime}\times2^{\prime}$ \\
G133.42+0.00 & 02:19:49.4 & +61:03:32 & -15.2 & 0.9$\pm$0.5 &  0.10 & $3^{\prime}\times3^{\prime}$ \\
G143.04+1.74 & 03:33:49.5 & +58:07:29 & -8.8 & 0.5 $^4$ &  0.14 & $3^{\prime}\times3^{\prime}$ \\
G154.05+5.07 & 04:47:09.0 & +53:03:26 & 4.6 & 0.2 $^5$ &  0.15 & $3^{\prime}\times3^{\prime}$ \\
  \hline\noalign{\smallskip}
\end{tabular}}
\end{center}
\begin{flushleft}
\footnotesize{ $^1$ The distance values of the sources are obtained from \citet{Immer+etal+2013}, $^2$ \citet{Ellsworth+etal+2015}, $^3$ \citet{Konyves+etal+2015}, $^4$ \citet{Stassun+etal+2018}, $^5$ \citet{Montillaud+etal+2015}, where the distances from \citet{Immer+etal+2013} and \citet{Konyves+etal+2015} are the parallax distances. The distance values of other sources are adopt the kinematic distance given in Paper II.}
\end{flushleft}
\end{table*}

Using the IRAM 30 m telescope, we searched for molecular spectral lines in 20 infall sources within the frequency ranges of approximately 83.7--91.5 GHz and 99.4--107.2 GHz. Table \ref{Tab:src} provides a list of these sources, including their coordinates, local standard of rest velocity ($v_{\text{LSR}}$) values, distances, the Root Mean Square (RMS) of the observations, and the extent of the observation maps. The coordinates and $v_{\text{LSR}}$ values are obtained from Paper II. For the distances of these sources, we prefer to use the values reported in the literature, especially those derived from the triangular parallax method, which are considered more reliable. If the distance cannot be found in the literature, we adopt the kinematic distance provided in Paper II. 

The observations were conducted in four projects with the following codes: 134-18 (February 20 to March 27, 2018), 032-19 (June 11 to June 19, 2019), 017-20 (November 11 to November 16, 2020), and 097-20 (January 27 to April 13, 2021), as shown in Table \ref{Tab:obs_par}. During these observations, the typical system temperature ranged from 90 to 135 K. We used the Eight Mixer Receivers (EMIR) as the receiver and the Fourier Transform Spectrometer (FTS) as the backend, with a bandwidth of 8 GHz and a frequency resolution of approximately 195 kHz. The on-the-fly position switch observing mode was used to cover each observation area. The angular resolution of the IRAM 30 m telescope ranges from 29$^{\prime\prime}$ at 83.7 GHz to 23$^{\prime\prime}$ at 107.2 GHz, with the main beam efficiency ranging from 81\% to more than 78\% across our observed frequency ranges. The size of the observation area was determined based on the MWISP $^{13}$CO (1-0) and C$^{18}$O (1-0) mapping results. Including the time required for source integration, pointing, focusing, and calibration, each source observation took approximately 1.5 to 3 hours.

The analysis of the HCO$^+$ (1-0) and H$^{13}$CO$^+$ (1-0) spectral lines for these sources has been provided in Papers III and IV. In this paper, we focus on the additional molecular spectral lines obtained from the FTS in the frequency ranges of 83.7 -- 91.5 GHz (lower side band) and 99.4 -- 107.2 GHz (upper side band). We used the CLASS software from the GILDAS package\footnote{\url{http://www.iram.fr/IRAMFR/GILDAS/}} to reduce the observed data. The median RMS values of these sources range from 0.10 to 0.22 K at a velocity resolution of about 0.67 km s$^{-1}$ (see Table \ref{Tab:src}). We then used the local thermodynamic equilibrium (LTE) radiation transfer model to identify and fit the observed spectral lines, and conducted further analysis on the physical properties of these sources.

\begin{table}
\begin{center}
  \caption{Observation parameters}\label{Tab:obs_par}  
 \setlength{\tabcolsep}{1mm}{
\hspace*{-1cm} 
\begin{tabular}{ccccccc}
  \hline\noalign{\smallskip}
Project & Time & Frequency & Detected Lines & RMS & Spectral & Angular  \\
ID &  & Coverage &  &  & Resolution &  Resolution \\
  \hline\noalign{\smallskip}
134-18 & Feb. 20 to Mar. 27, 2018 & \multirow{2}{*}{83.7--91.5 GHz} & HCN 1-0, HCO$^+$ 1-0, HNC 1-0, & \multirow{2}{*}{0.10--0.22 K} & \multirow{2}{*}{195 kHz} & \multirow{2}{*}{29$^{\prime\prime}$} \\
032-19 & Jun. 11 to 19, 2019 &  & c-C$_3$H$_2$ 2(1,2)-1(0,1), CCH 1-0, et al. &  &  & \\
017-20 & Nov. 11 to 16, 2020 & \multirow{2}{*}{99.4--107.2 GHz} & HC$_3$N 11-10, H$_2$CS 3-2, SO N=5-4 J=4-4, &  \multirow{2}{*}{0.10--0.27 K}  & \multirow{2}{*}{195 kHz} & \multirow{2}{*}{23$^{\prime\prime}$} \\
097-20 & Jan. 27 to Apr. 13, 2021 &  & SO$_2$ 3(1,3)-2(0,2), et al. &  &  & \\
  \hline\noalign{\smallskip}
\end{tabular}}
\end{center}
\end{table}

\section{Results}\label{sec:result}
In previous works, we identified 29 sources with gas infall motion (Papers III and IV). These sources show blue profiles in the optically thick line HCO$^+$ 1-0. In some cases, weak self-absorption or large shifts in the spectral profile make the red peak of the blue profile less obvious. This results in a peak-shoulder profile or a single-peaked profile with the peak skewed to the blue \citep[e.g.][]{Klaassen+etal+2007, Wyrowski+etal+2016, Devine+etal+2018}. Among these sources, 15 display typical blue profiles, eight show peak-shoulder profiles, and the remaining six have single-peaked profile with the peak skewed to the blue. In Paper V, we analyzed nine sources with typical blue profiles. Here, we present the results for the remaining 20 sources.

\subsection{Line Identification}\label{subsec:lines}

\begin{table*}
\begin{center}
  \caption{List of observed molecular lines}\label{Tab:freq}  
 \setlength{\tabcolsep}{0.1mm}{
\begin{tabular}{ccccc @{\vline} ccccc}

  \hline\noalign{\smallskip}
Molecular & Transition & Frequency & $E_u$ & $A_{ul}$ & Molecular & Transition & Frequency & $E_u$ & $A_{ul}$ \\
 species &  &  (MHz) & (K) & (s$^{-1}$) & species  &  &  (MHz) & (K) & (s$^{-1}$) \\
  \hline\noalign{\smallskip}
CH$_3$OH  &  J=5(-1,5)-4(0,4) E  &84521.2 & 40.4 & 2.0$\times$10$^{-6}$ &  H$^{13}$CO$^+$  &  J=1-0  &86754.3  & 4.2 & 3.9$\times$10$^{-5}$ \\
c-C$_3$H$_2$  &  J=2(1,2)-1(0,1)  &85338.9 & 6.4 & 2.3$\times$10$^{-5}$ & SiO  &  J=2-1  &86847.0 & 6.3 & 2.9$\times$10$^{-5}$ \\
& J=4(3,2)-4(2,3) & 85656.4 & 29.1 & 1.5$\times$10$^{-5}$ &  HN$^{13}$C  & J=1-0 & 87090.9 & 4.2 & 2.4$\times$10$^{-5}$ \\
HCS$^+$  &  J=2-1  &85347.9 & 6.1 & 1.1$\times$10$^{-5}$ &  CCH  &  N=1-0 J=3/2-1/2 F=1-1  &87284.2 & 4.2 & 2.8$\times$10$^{-7}$ \\
CH$_3$CCH  &  J=5(3)-4(3)  &85442.6 & 77.3 & 1.2$\times$10$^{-6}$ &    &  N=1-0 J=3/2-1/2 F=2-1  &87316.9 & 4.2 & 1.5$\times$10$^{-6}$  \\
  &  J=5(2)-4(2)  &85450.8 & 41.2 & 1.6$\times$10$^{-6}$ &   &  N=1-0 J=3/2-1/2 F=1-0  &87328.6 & 4.2 & 1.3$\times$10$^{-6}$  \\
  &  J=5(1)-4(1)  &85455.7 & 19.5 &  1.8$\times$10$^{-6}$ &   &  N=1-0 J=1/2-1/2 F=1-1  &87402.0 & 4.2 & 1.3$\times$10$^{-6}$ \\
   &  J=5(0)-4(0)  &85457.3 & 12.3 & 1.9$\times$10$^{-6}$ &   &  N=1-0 J=1/2-1/2 F=0-1  &87407.2 & 4.2 & 1.5$\times$10$^{-6}$ \\
  &  J=6(4)-5(4)  &102516.6 & 132.8 & 1.8$\times$10$^{-6}$ &    &  N=1-0 J=1/2-1/2 F=1-0  &87446.5 & 4.2 & 2.6$\times$10$^{-7}$ \\
  &  J=6(3)-5(3)  &102530.3 &  82.3 & 2.4$\times$10$^{-6}$ &  HNCO  &  J=4(0,4)-3(0,3)  &87925.2 & 10.5 & 8.8$\times$10$^{-6}$ \\
  &  J=6(2)-5(2)  &102540.1 & 46.1 & 2.9$\times$10$^{-6}$ &   & J=4(1,3)-3(1,2) &88239.0 & 53.9 & 8.2$\times$10$^{-6}$ \\
  &  J=6(1)-5(1)  &102546.0 & 24.5 & 3.2$\times$10$^{-6}$ &  HCN  &  J=1-0 F=1-1  &88630.4 & 4.3 & 2.4$\times$10$^{-5}$ \\
  &  J=6(0)-5(0)  &102548.0 & 17.2 & 3.3$\times$10$^{-6}$ &    &  J=1-0 F=2-1  &88631.8 & 4.3 & 2.4$\times$10$^{-5}$ \\
C$_4$H  &  N=9-8 J=19/2-17/2  &85634.0 & 20.5 & 2.6$\times$10$^{-6}$ &   &  J=1-0 F=0-1  &88633.9 & 4.3  & 2.4$\times$10$^{-5}$ \\
  &  N=9-8 J=17/2-15/2  &85672.6 & 20.6 & 2.6$\times$10$^{-6}$ &  HCO$^+$  &  J=1-0  &89188.5 & 4.3 & 4.2$\times$10$^{-5}$ \\
H$\alpha$ & H(42)$\alpha$ & 85688.4 &  &  &  HNC  &  J=1-0 &90663.6 & 4.4 & 2.7$\times$10$^{-5}$ \\
NH$_2$D  &  J=1(1,1)0+ -1(0,1)0- F=2-2  &85926.3 & 20.7 & 5.9$\times$10$^{-6}$ &  HC$_3$N  &  J=10-9  &90979.0 & 24.0 & 5.8$\times$10$^{-5}$ \\
HC$^{15}$N  &  J=1-0  & 86055.0 & 4.1 & 2.2$\times$10$^{-5}$ &    &  J=11-10  &100076.4 & 28.8 & 7.8$\times$10$^{-5}$ \\
SO  &  N=2-1 J=2-1  & 86094.0 & 19.3 & 5.4$\times$10$^{-6}$ &  H$_2$CS  &  J=3(1,3)-2(1,2)  &101477.9 & 22.9 & 1.3$\times$10$^{-5}$ \\
& N=5-4 J=4-4 & 100029.6 & 38.6 & 1.1$\times$10$^{-6}$ &    &  J=3(0,3)-2(0,2)  &103040.5 & 9.9 & 1.5$\times$10$^{-5}$ \\
H$^{13}$CN  &  J=1-0 F=1-1  &86338.7 & 4.1 & 2.2$\times$10$^{-5}$ &    &  J=3(1,2)-2(1,1)  &104617.0 & 23.2 & 1.4$\times$10$^{-5}$ \\
  &  J=1-0 F=2-1  &86340.2 & 4.1 & 2.2$\times$10$^{-5}$ &  SO$_2$  &  J=3(1,3)-2(0,2)  &104029.4 & 7.7 & 1.0$\times$10$^{-5}$ \\
  &  J=1-0 F=0-1  & 86342.3 & 4.1 & 2.2$\times$10$^{-5}$ &  &  J=10(1,9)-10(0,10) &104239.3 & 54.7 & 1.1$\times$10$^{-5}$ \\
HCO  &  N=1(0,1)-0(0,0) J=3/2-1/2 F=2-1  &86670.8  & 4.2 & 4.7$\times$10$^{-6}$ & $^{13}$C$^{18}$O & J=1-0 & 104711.4 & 5.0 & 5.5$\times$10$^{-8}$ \\
  &  N=1(0,1)-0(0,0) J=3/2-1/2 F=1-0  &86708.4 & 4.2 & 4.6$\times$10$^{-6}$ &  &  &&&\\
  &  N=1(0,1)-0(0,0) J=1/2-1/2 F=1-1  &86777.4 & 4.2 & 4.6$\times$10$^{-6}$ &  &  &&&\\
  &  N=1(0,1)-0(0,0) J=1/2-1/2 F=0-1  &86805.8 & 4.2 & 4.7$\times$10$^{-6}$ &  &  &&&\\

  \hline\noalign{\smallskip}
\end{tabular}}
\end{center}
\footnotesize{The spectroscopic data are from \url{https://splatalogue.online//}.}
\end{table*}

\begin{figure*}
\hspace*{-1.cm}
\includegraphics[width=1.1\textwidth]{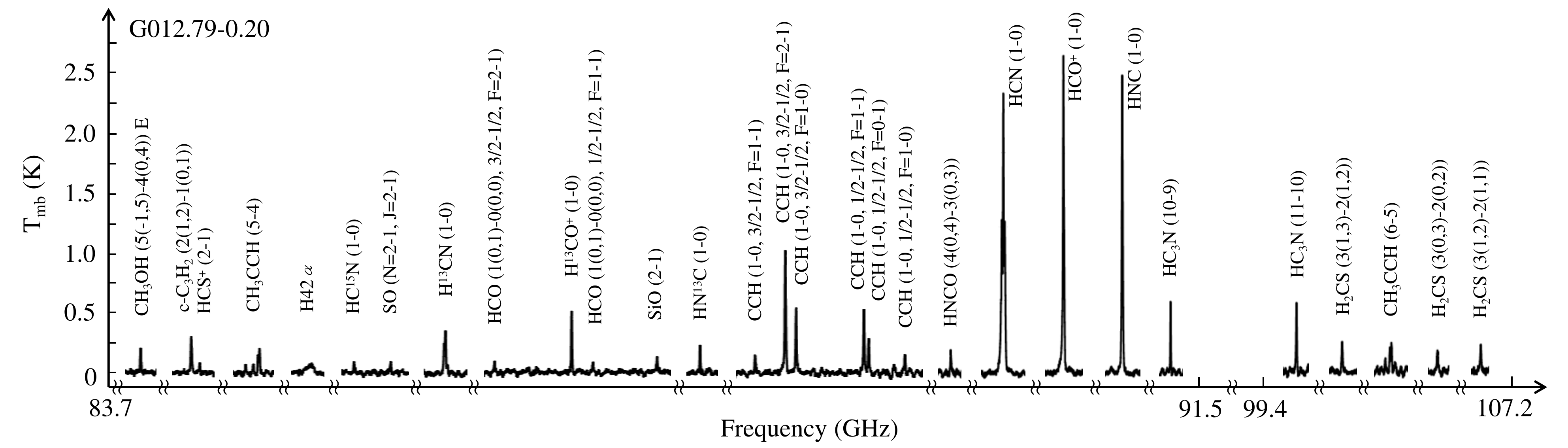}
\hfill
\caption{The spectrum of the molecular emission lines in G012.79-0.20, observed using FTS with two bands covering 8 GHz and a frequency resolution of 195 kHz.}
\label{fig:lines}
\end{figure*}

Multiple molecular emission lines were detected in these 20 sources. The observed spectral lines are identified using XCLASS \footnote{\url{https://xclass.astro.uni-koeln.de/Home}} \citep{Moller+etal+2017}, and the spectroscopic data are taken from the Cologne Database of Molecular Spectroscopy \citep[CDMS;][]{Muller+etal+2001, Endres+etal+2016} and SPLATALOGUE \footnote{\url{https://splatalogue.online//}}. After excluding the bad channels, the spectral lines in the averaged spectra from the observed region of each source with a signal-to-noise ratio (SNR) greater than 3 are selected for further analysis. To identify the detected molecular emission lines, we searched for transition lines within the observed frequency range and with energies of the upper level less than 1000 K. We simulate the emission spectrum of a molecular species. Considering the previous observational results, where the kinetic temperatures of our sources are mostly around 20 K, the column densities of H$^{13}$CO$^+$ range from $10^{12}$ to $10^{13}$ cm$^{-2}$, the column densities of HCO$^+$ range from $10^{13}$ to $10^{14}$ cm$^{-2}$, and the mean line width of H$^{13}$CO$^+$ is approximately 2 km s$^{-1}$, we assume an excitation temperature of 20 K, a column density of $1 \times 10^{13}$ cm$^{-2}$, and a line width of approximately 2 km s$^{-1}$ for the molecular species. The assumed central radial velocity of the molecular emission spectrum is determined based on the central radial velocity of H$^{13}$CO$^+$ or C$^{18}$O, which is observed in the source. If the LTE simulated spectrum of a species roughly matches the observed spectral lines, that species is identified as being present (more detailed fitting for each clump is done in a following step, see details in section \ref{subsec:para}). Figure \ref{fig:lines} shows the spectrum of G012.79-0.20 as an example (the complete spectra of all sources are provided online \footnote{\url{https://github.com/yangy4068/2024/tree/Figure-1}}). The figure displays a segment of the spectrum analyzer band, divided into multiple segments with molecular lines, within the observed frequency range. The molecular species names and transition energy levels corresponding to each emission line are indicated in the figure. We list the recognized molecular species names, transition energy levels, and rest frequencies for these lines in Table \ref{Tab:freq}.

By using the aforementioned method, emission lines of up to 22 different molecular species (including a few isotopologues) and one hydrogen radio recombination line are identified in 20 observed sources. Among them, 15 sources show H$^{13}$CO$^+$, HCO$^+$, HCN, HNC, c-C$_3$H$_2$, and CCH lines. G012.79-0.20 exhibits the largest number of distinct molecular emission lines. In this source, more than 30 transition lines from 18 molecular species and one atomic line are detected. G110.40+1.67 has the fewest detected lines. In the averaged spectrum over the entire observed region ($2.5^{\prime}\times2.5^{\prime}$) of G110.40+1.67, only three lines, HCN 1-0, HCO$^+$ 1-0, and HNC 1-0, are identified. For most of the observed sources, multiple transition lines of certain molecular species, such as CCH, CH$_3$CCH, HCO, HC$_3$N, and H$_2$CS, are identified within the observed frequency range. We can further constrain the physical parameters, such as excitation temperature and column density, through fitting these multiple transition lines. In addition, a second velocity component is identified in G012.96-0.23, G025.82-0.18, G049.07-0.33, and G133.42+0.00, corresponding to velocities of approximately 52.5, 113.0, 67.4, and -46.0 km s$^{-1}$, respectively. In our subsequent analysis, we will also separate and discuss these different velocity components individually. 

\subsection{Mapping Results} \label{subsec:map}

\begin{figure*}
\hspace*{-1.cm}
\includegraphics[width=1.05\textwidth]{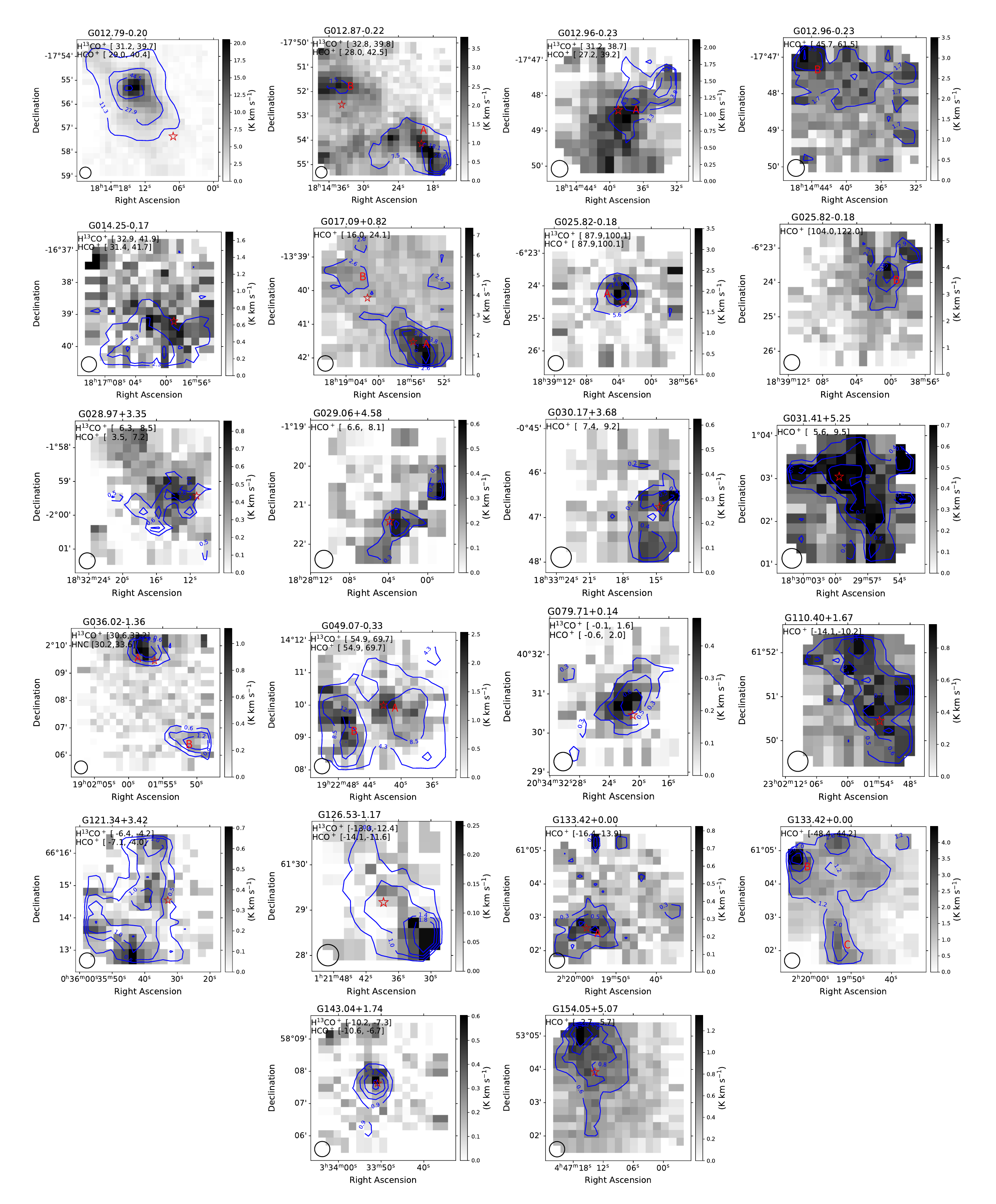}
\hfill
\caption{The HCO$^{+}$ (or HNC) integrated intensity contour (blue) maps for 20 sources are superimposed on their H$^{13}$CO$^{+}$ integrated intensity images. If H$^{13}$CO$^{+}$ emission is not detected, only the HCO$^{+}$ map is provided. The contours in the panels start at 3$\sigma$ (for G012.79-0.20, G012.87-0.22 and G012.96-0.23, start at 10$\sigma$). The molecular name and the integrated velocity range are marked in the top left corner of the image. The pentagram symbol denotes the position where the most prominent infall profile was identified in Papers III and IV. For sources that have more than one clump within the observed areas, we have labeled these clumps as A, B and C.}
\label{fig:all_maps}
\end{figure*}

\begin{figure*}
  \begin{minipage}[t]{0.33\linewidth}
  \centering
	\includegraphics[width=2in]{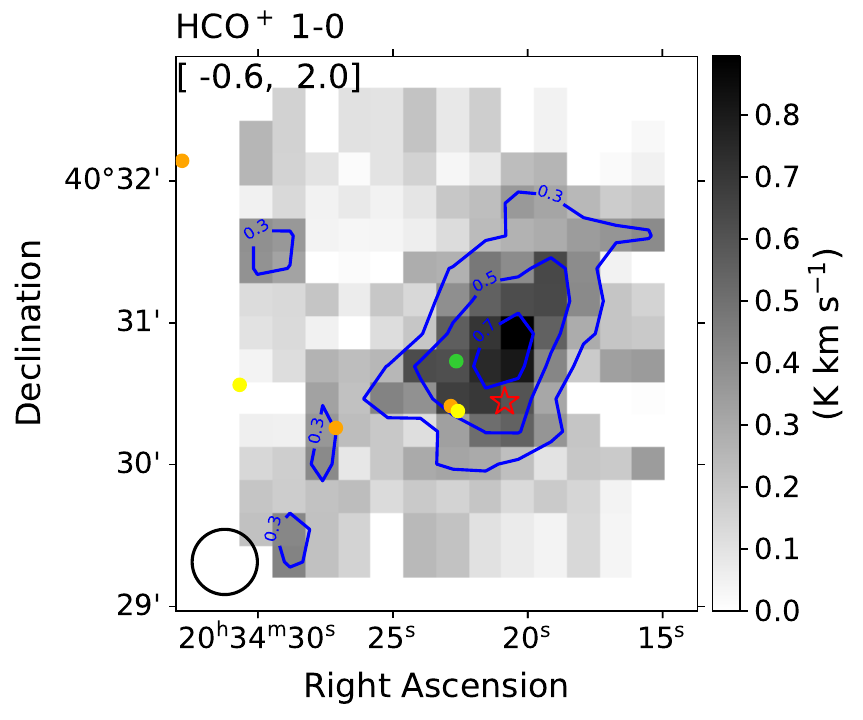}
  \end{minipage}%
  \begin{minipage}[t]{0.33\linewidth}
  \centering
	\includegraphics[width=2in]{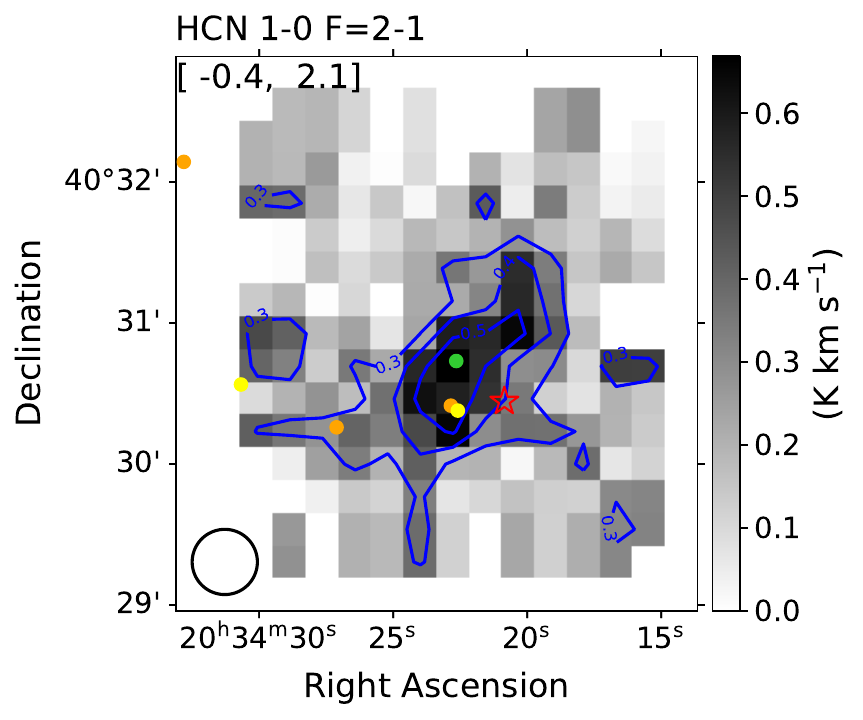}
  \end{minipage}%
  \begin{minipage}[t]{0.33\linewidth}
  \centering
	\includegraphics[width=2in]{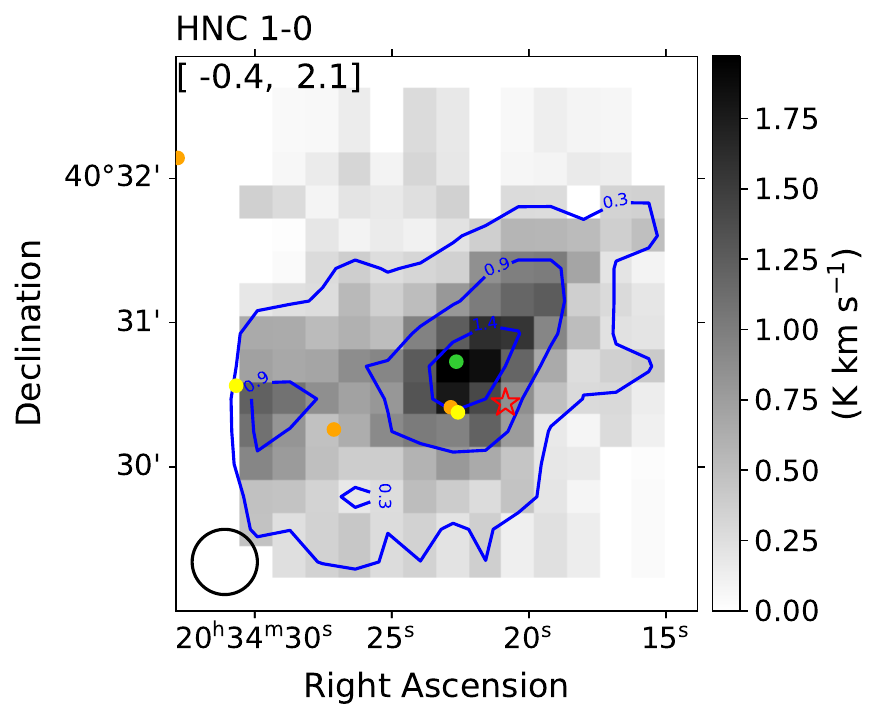}
  \end{minipage}%
  
  \begin{minipage}[t]{0.33\linewidth}
  \centering
	\includegraphics[width=2in]{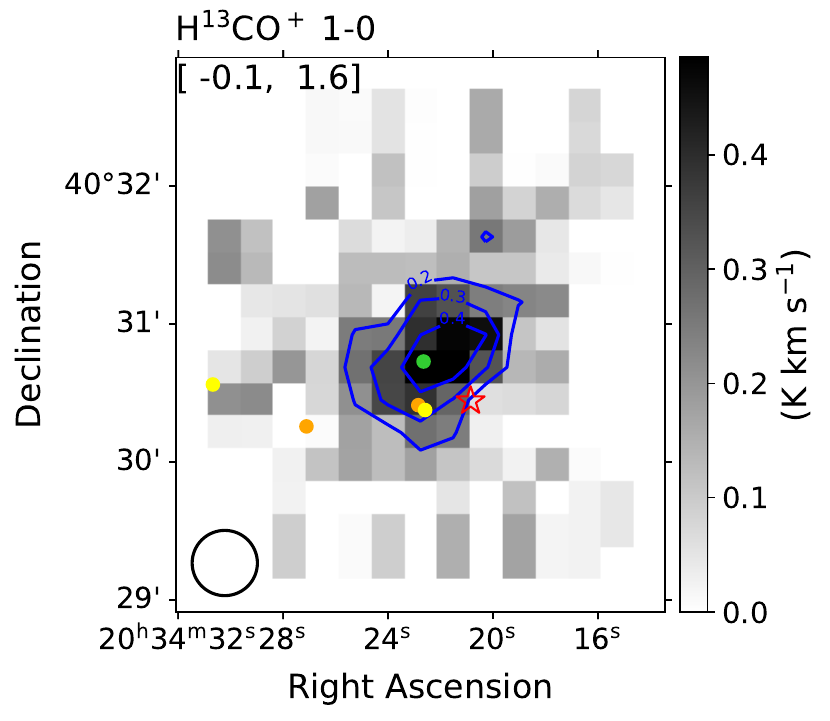}
  \end{minipage}%
  \begin{minipage}[t]{0.33\linewidth}
  \centering
	\includegraphics[width=2in]{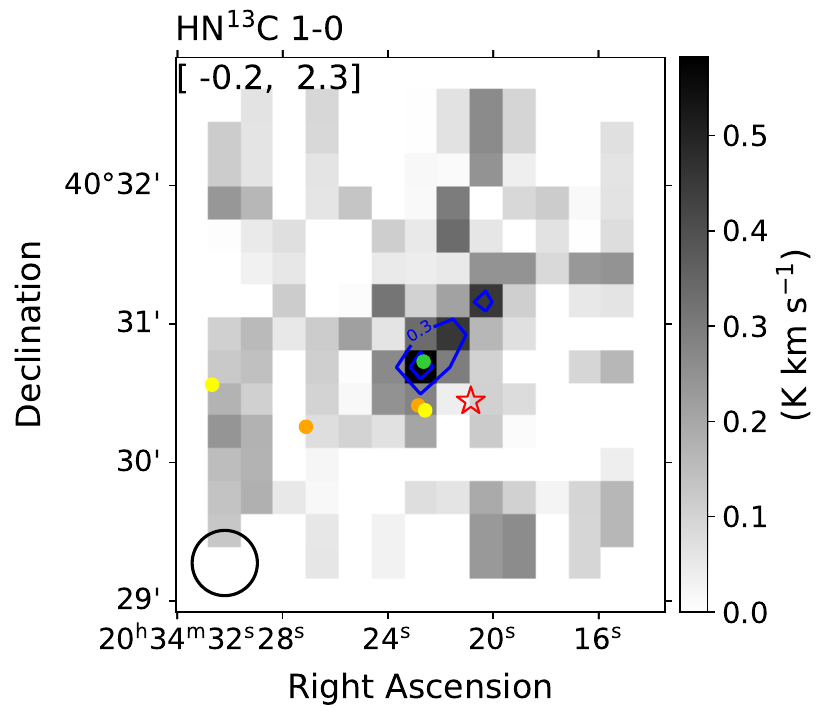}
  \end{minipage}%
  \begin{minipage}[t]{0.33\linewidth}
  \centering
	\includegraphics[width=2in]{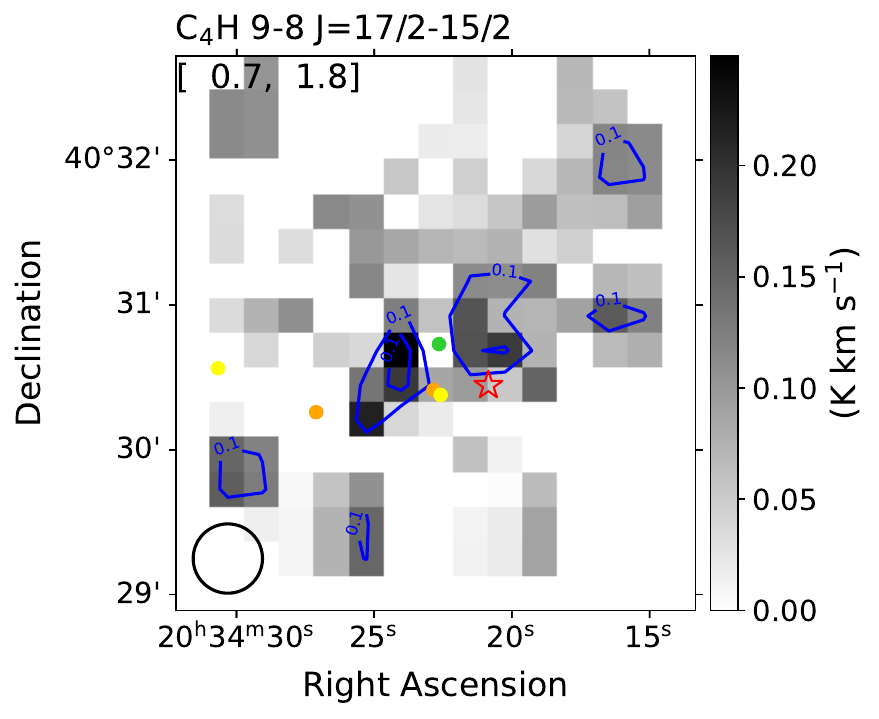}
  \end{minipage}%
  
  \begin{minipage}[t]{0.33\linewidth}
  \centering
	\includegraphics[width=2in]{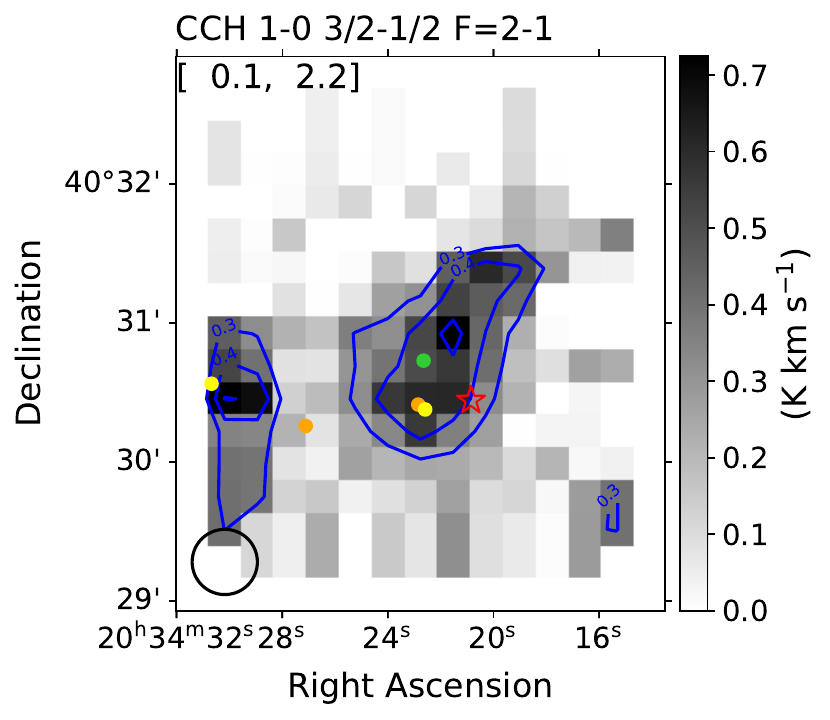}
  \end{minipage}%
  \begin{minipage}[t]{0.33\linewidth}
  \centering
	\includegraphics[width=2in]{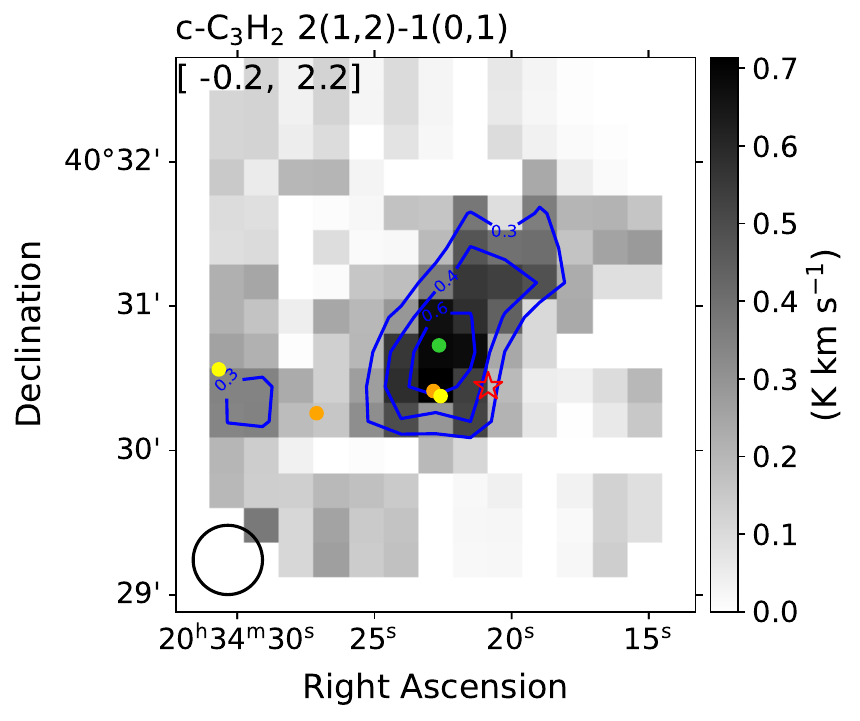}
  \end{minipage}%
  \begin{minipage}[t]{0.33\linewidth}
  \centering
	\includegraphics[width=2in]{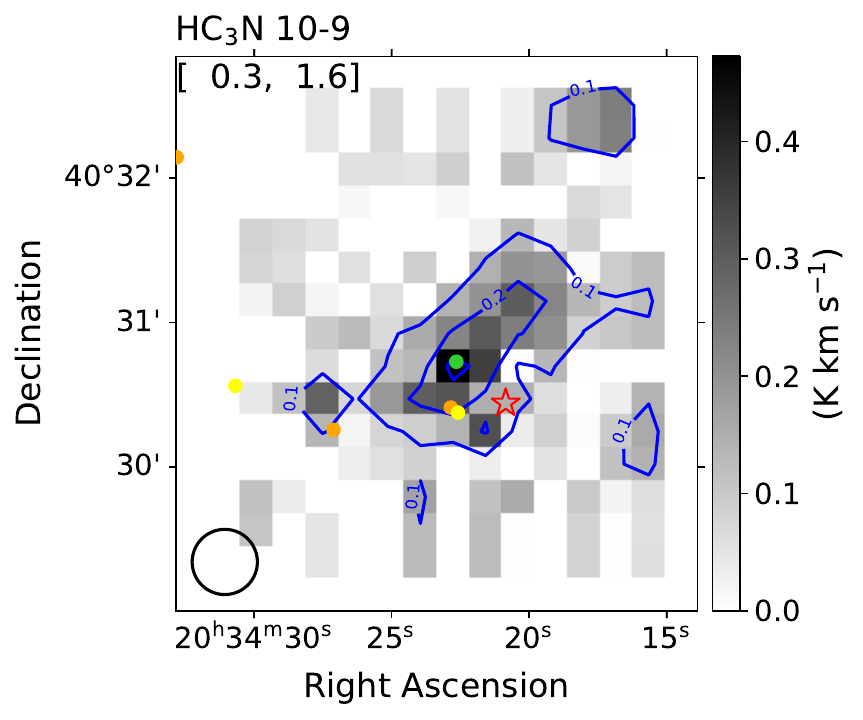}
  \end{minipage}%

\caption{Example of source G079.71+0.14: integrated intensity maps of nine molecular species. The contours (blue) are the [0.2, 0.4, 0.6, 0.8] of the maximum value of its integrated intensity. The pentagram symbol denotes the position where the most prominent infall profile was identified in Paper III. The green, yellow, and orange points denote the Class I, Flat-spectrum, Class II YSOs, respectively. The black circle at the bottom left represents the beam size of the IRAM 30 m telescope.}
\label{fig:maps}
\end{figure*}

Figure \ref{fig:all_maps} displays the HCO$^{+}$ (or HNC) integrated intensity maps for 20 infall sources. The integrated intensity contours are superimposed on the H$^{13}$CO$^{+}$ integrated intensity images. For sources with multiple velocity components, we integrate each velocity component separately. For sources that have more than one clump within the observed areas, we have labeled these clumps as A, B and C. In Figure \ref{fig:maps}, we present integrated intensity maps for all molecular species observed in our source (G079.71+0.14 is shown here as an example). The complete figure set of all sources is provided online \footnote{\url{https://github.com/yangy4068/2024/tree/Figure-3}}. For molecules with multiple transition lines, we only display the integrated intensity map of the line with the highest SNR. The molecular name, transition energy level, and the corresponding integrated velocity range are marked at the upper-left corner of the figures. In addition, the positions of young stellar objects (YSOs), and masers are also displayed on the maps. The YSOs are identified in the Spitzer/IRAC candidate YSOs (SPICY) catalog \citep{Kuhn+etal+2021}. If the coordinates of our sources are outside the coverage range of the SPICY catalog, we use the AllWISE sources \citep{vizier:II328} and classify them based on the YSOs criteria suggested by \citet{Koenig+etal+2012}.  The positions of the masers are sourced from \citet{Yang+etal+2017, Yang+etal+2019, Valdettaro+etal+2001, Anglada+etal+1996, Qiao+etal+2016, Qiao+etal+2018, Qiao+etal+2020}, marked with crosses of different colors: blue for 95 GHz methanol maser, magenta for 6.7 GHz methanol maser, orange for H$_2$O maser, and red for OH maser. The velocities of the masers we adopted are close to the $V\rm_{LSR}$ of the observed sources. The red pentagrams denote the positions of the most significant infall profile of HCO$^+$ (1-0) lines identified in Papers III and IV.

As can be seen from these figures, in most cases, the integrated intensity maps of various molecular species show spatially similar compact regions. However, in some cases, the spatial distributions of certain molecules species lack a clumpy structure, such as G028.97+3.35, G030.17+3.68, G110.40+1.67 and G121.34+3.42. For these sources, we only analyze their average spectra within the observation area in the next section. For other sources that exhibit compact structures, we introduce them individually below.

In \textbf{G014.25-0.17, G017.09+0.82, G029.06+4.58, G031.41+5.25, G037.05-0.03, G079.71+0.14, G126.53-1.17, G143.04+1.74, and G154.05+5.07}, the spatial distributions of various molecular lines are quite similar, all concentrated in the central region of the clumps where gas infall has been observed. 

\textbf{G012.79-0.20, G012.87-0.22 and G012.96-0.23}: The clump in G012.79-0.20 is associated with the W33 Main region. According to Paper IV, the HCO$^+$ lines in the central region of this clump show three-peaked profiles, which suggest complex internal structures. The confirmed gas infall areas are mainly located in the southwestern of the clump. Our observations have detected the emissions of the hydrogen radio recombination line (RRL) H(42)$\alpha$ in the central region of the clump. Additionally, we have detected several carbon chain molecules, such as CH$_3$CCH and HC$_3$N. The detected SiO in this clump may be associated with gas outflow activity. G012.87-0.22 is located in the northeast area of the W33 Main region. Our observations of HCN, HCO$^+$ and HNC have revealed two clumps in the target area. However, the integrated intensity maps of HC$_3$N and CCH show a third clump, which coincides with an extension of the HCN, HCO$^+$, and HNC emissions. The isotopologues H$^{13}$CO$^+$ and HN$^{13}$C also display similar structures, though these structural features are less pronounced. Weak SiO emissions have also been detected in the average spectra of this target source, but due to a limited SNR, no discernible centrally peaked structure have been observed. In the northern area of the W33 A region, the source G012.96-0.23 displays two close velocity components with velocity ranges of [27.2, 41.0] and [45.1, 61.5] km s$^{-1}$, respectively. The first one is associated with observed gas infall motions, while the second one, only showing emissions of HCO$^+$, HCN, and HNC, is located in the northeast part of the observation area, slightly offset from the area where gas infall occurs. These three sources have more kinds of species detected. This may be related to the higher SNR in these sources and their relatively high H$_2$ column densities and kinetic temperatures.

\textbf{G025.82-0.18:} This source shows two velocity components, with the component at 93.7 km s$^{-1}$ associated with gas infall motion (clump A) and the other velocity component at 113.0 km s$^{-1}$, displaying clump B in the integrated intensity maps, with no gas infall detected.

\textbf{G036.02-1.36:} In G036.02-1.36, maps of several spectral lines, such as HNC and c-C$_3$H$_2$, reveal two clumps within the observation area, labeled as Clump A and B, with Clump A showing evidence of infall motion. 

\textbf{G049.07-0.33:} This source shows two close velocity components, spatially separated in the integrated intensity map: the velocity component [54.9,62.6] km s$^{-1}$ is associated with the confirmed infall source, while the [62.6,74.9] km s$^{-1}$ component traces another molecular clump which is located in the eastern part of the observation area. 

\textbf{G133.42+0.00:} This source shows two velocity components, with the component at -15.2 km s$^{-1}$ associated with gas infall motion (clump A) and another velocity component located at approximately -46 km s$^{-1}$, displaying two clump structures (clump B and C) in the integrated intensity maps.

\section{Analysis and Discussion} \label{sec:analysis}

\subsection{Parameters of the Clumps}\label{subsec:size}

\begin{table*}
\caption{Parameters of the Clumps}\label{Tab:para}
\setlength{\tabcolsep}{0.1mm}{
\hspace*{-1cm} 
\begin{tabular}{ccccccccccccc}

  \hline\noalign{\smallskip}
Clump & \multicolumn{2}{c}{Center Coordinate $^1$} & \multirow{2}{*}{$v_{\text{LSR}}^2$} & \multirow{2}{*}{Radius} & Aspect & \multirow{2}{*}{$T_{\text{kin}} ^4$} & \multirow{2}{*}{log($\frac{N({\rm H}_2)}{{\rm cm}^{-2}}$)$^4$} & \multirow{2}{*}{log($\frac{M}{M_{\odot}}$)$^5$} & \multirow{2}{*}{$ v_{in} ^6$} & $\dot  M_{in} ^6$ & Line & Number of \\
Name & R.A. & Decl. & &  & Ratio$^3$ & & &  &  & $\times10^{-4}$ & Profile$^7$  & Molecules$^8$ \\ 
 & (J2000) & (J2000) & (km s$^{-1}$) & (pc) &  & (K) & &  & ($\mathrm{km\,s}^{-1}$) & ($ M_{\odot}\,\mathrm{yr}^{-1}$) &  & \\ 
  \hline\noalign{\smallskip}
G012.79-0.20 & 18:14:14.5 & -17:55:19 & 35.8 & 0.42(0.07) & 1.28 & 33.0 & 23.1 & 3.0 & 0.44--1.45 & ... & B & 19\\ 
G012.87-0.22 A & 18:14:18.4 & -17:54:37 & 35.4 & 0.36(0.15) & 3.38 & 22.3 & 23.2 & 3.0 & 0.06--1.37 & ... & B & 15\\ 
G012.87-0.22 B & 18:14:34.6 & -17:51:49 & 35.4 & 0.32(0.23) & 1.27 & 25.5 & 23.0 & 2.7 & 0.24$\pm$0.01 & 12& B & 15\\ 
G012.96-0.23 A & 18:14:38.0 & -17:48:41 & 35.2 & 0.25(0.17) & 2.23 & 17.1 & 23.1 & 2.6 & 1.46$\pm$0.11 & 66 & B & 8\\ 
G012.96-0.23 B$^9$ & 18:14:44.9 & -17.49:52 & 52.5 & 0.21(0.08) & 2.52 & 17.1 & 23.1 & 2.4 & ... & ... & N & 3 \\ 
G014.25-0.17 & 18:16:58.2 & -16:39:23 & 38.1 & 0.54(0.28) & 1.99 & 20.9 & 22.9 & 3.1 & 0.94$\pm$0.40 & 55 & B & 6\\ 
G017.09+0.82 A$^9$ & 18:18:54.8 & -13:41:45 & 22.3 & 0.32(0.23) & 2.44 & 25.0 & 22.6 & 2.3 & 0.45$\pm$0.11 & 8.7 & B & 5\\ 
G017.09+0.82 B$^9$ & 18:19:03.6 & -13:39:38 & 22.3 & 0.16(0.06) & 1.05 & 25.0 & 22.6 & 1.7 & 0.56$\pm$0.08 & 5.4 & B & 5\\ 
G025.82-0.18 A & 18:39:04.2 & -06:24:15 & 93.7 & 0.53(0.12) & 1.02 & 17.4 & 22.8 & 3.0 & 1.85$\pm$0.53 & 85 & B & 6\\ 
G025.82-0.18 B$^9$ & 18:39:00.4 & -06:23:33 & 113.0  & 0.57(0.24) & 1.04 & ... & ... & ...& ... & ... & N & 3\\ 
G028.97+3.35$^{10}$ & 18:32:13.7 & -01:59:28 & 7.3 & ... & ... & 12.4 & 22.4 & ... & ... & ... & S & 10\\ 
G029.06+4.58$^9$ & 18:28:03.3 & -01:21:32 & 7.5 & 0.05(0.01) & 1.02 & 8.7 & 21.6 & $\textless$ 1 & ... & ... & S & 5\\ 
G030.17+3.68$^{10}$ & 18:33:14.4 & -00:46:44 & 9.0 & ... & ... & 10.4 & 21.8 & ... & ... & ... & P-S & 5\\ 
G031.41+5.25$^9$ & 18:29:58.7 & +01:03:09 & 8.3 & 0.07(0.06) & 1.44 & 13.8 & 22.0 & $\textless$ 1 & ... & ... & P-S & 5\\ 
G036.02-1.36 A & 19:01:58.4 & +02:09:48 & 31.7 & 0.20(0.09) & 1.19 & 10.0 & 21.9 & 1.2 & ... & ... & P-S & 6\\ 
G036.02-1.36 B$^9$ & 19:01:50.0 & +02:06:30 & 31.7 & 0.28(0.09) & 1.33 & 10.0 & 21.9 & 1.5 & ... & ... & N & 3\\ 
G037.05-0.03 & 18:59:04.4 & +03:38:27 & 81.4 & 0.36(0.14) & 2.26 & 13.2 & 22.5 & 2.3 & 1.58$\pm$0.40 & 28 & P-S & 6\\ 
G049.07-0.33 A & 19:22:42.3 & +14:10:00 & 60.9 & 0.51(0.22) & 1.28 & 19.5 & 22.5 & 2.6 & ... & ... & P-S & 6\\ 
G049.07-0.33 B & 19:22:47.2 & +14:09:30 & 67.4 & 0.46(0.19) & 1.49 & 19.5 & 22.5 & 2.5 & ... & ... & N & 6\\ 
G079.71+0.14 & 20:34:22.7 & +40:30:40 & 1.2 & 0.07(0.02) & 1.27 & 9.1 & 21.7 & $\textless$ 1 & ... & ... & P-S & 9\\ 
G110.40+1.67$^{10}$ & 23:01:57.9 & +61:50:55 & -11.2 & ... & ... & 12.6 & 21.8 & ... & ... & ... & P-S & 3\\ 
G121.34+3.42$^{10}$ & 00:35:39.0 & +66:14:34 & -5.2 & ... & ... & 7.2 & 22.0 & ... & ... & ... & S & 6\\ 
G126.53-1.17 & 01:21:31:6 & +61:28:18 & -12.3 & 0.05(0.03) & 1.20 & 11.9 & 22.0 & $\textless$ 1 & ... & ... & S & 6\\ 
G133.42+0.00 A$^9$ & 02:19:55.0 & +61:02:36 & -15.2 & 0.07(0.04) & 1.38 & 10.6 & 21.7 & $\textless$ 1 & 0.47$\pm$0.19 & 0.24 & P-S & 5\\ 
G133.42+0.00 B$^9$ & 02:20:04.5 & +61:04:54 & -46.0  & 0.29(0.09) & 1.96 & ... & ... & ... & ... & ... & N & 3\\ 
G133.42+0.00 C$^9$ & 02:19:52.5 & +61:02:13 & -46.0  & 0.21(0.05) & 2.10 & ... & ... & ... & ... & ... & N & 4\\ 
G143.04+1.74 & 03:33:51.3 & +58:07:49 & -8.8 & 0.06(0.01) & 1.21 & 14.6 & 22.0 & $\textless$ 1 & ... & ... & S & 6\\ 
G154.05+5.07$^9$ & 04:47:16.7 & +53:04:49 & 4.6 & 0.05(0.01) & 1.33 & 10.3 & 21.5 & $\textless$ 1 & ... & ... & S & 4\\ 
  \hline\noalign{\smallskip}
\end{tabular}}
\footnotesize{$^1$ The center coordinates of the clump are the center of the two-dimensional Gaussian fitting of the integrated intensity map (for sources without discernible clumpy structures, the coordinates are the center of the observation area). $^2$ The $v_{\text{LSR}}$ of the clump is usually traced by the optically thin line H$^{13}$CO$^+$. $^3$ The aspect ratio of the clump is obtained from the two-dimensional Gaussian fitting of the integrated intensity map. $^4$ The values of $T_{\text{kin}}$ and $N(\text{H}_2)$ for these sources are obtained from Paper II, where $T{\text{kin}}$ is derived from $^{12}$CO data, and $N(\text{H}_2)$ is estimated through C$^{18}$O data. $^5$ The mass values are obtained from Papers III and IV (estimated from the H$2$ column density). $^6$ The infall velocity and mass infall rate values are adopted from Papers III and IV (obtained by fitting the line profiles of HCO$^+$ and H$^{13}$CO$^+$). $^7$ B denotes a double-peaked blue profile in HCO$^+$ (1-0), P-S denotes a peak-shoulder profile, S denotes a single-peaked profile with the peak skewed to the blue, and N denotes a symmetric line profile or multi-peaked profile. $^8$ The number of identified molecular species within the clump area. $^9$ Since H$^{13}$CO$^+$ was not detected or was too weak in the sources, HCO$^+$ or HNC was used to obtain the clump parameters, and Gaussian fitting was used to trace the clump's $v_{\text{LSR}}$. $^{10}$ Source with no discernible clumpy structure.
}
\end{table*}

We distinguish molecular clumps based on their velocity components and spatial positions. The sizes of molecular clumps are estimated using two-dimensional Gaussian fitting on integrated intensity maps of H$^{13}$CO$^+$. We chose H$^{13}$CO$^+$ because it is generally optically thin and does not have complex velocity components, facilitating clear identification of clumps. If H$^{13}$CO$^+$ emission is not detected in a source, we use HCO$^+$ or HNC data to identify clumps  (this happens for a total of 10 clumps, which are marked in Table \ref{Tab:para}). For sources showing clumpy structures, the center coordinate of the clump is the center of the two-dimensional Gaussian fitting. The radius ($R$) is calculated as the geometric mean of the FWHM of the semi-major axis ($a$) and semi-minor axis ($b$), using $R = \sqrt{a \times b}$, and subsequently converted from angular to linear sizes. For sources without clear clumpy structures (a total of four sources, which are marked in Table \ref{Tab:para}), the central coordinate is the center of the observation area, as our target area is also determined by the clump identified from $^{13}$CO and/or C$^{18}$O data. All these central coordinates, radius, and aspect ratio (i.e., $a/b$) of the clumps are listed in Table \ref{Tab:para}.

For the sources whose spatial positions and velocity components are consistent with those in Paper II, we adopted the kinetic temperature ($T_{\text{kin}}$) and H$_2$ column density data provided by Paper II. For the sources with consistent velocity components but inconsistent spatial positions, we use the same method as in Paper II to estimate their temperature and column density. Among them, $T_{\text{kin}}$ is derived from the CO excitation temperature of MWISP data, with an uncertainty of less than 20\%. The H$_2$ column densities are estimated through C$^{18}$O data, assuming a column density ratio of H$_2$ to C$^{18}$O as $N(\text{H}_2)/N(\text{C}^{18}\text{O})=7\times10^6$ \citep[e.g.,][]{Castets+langer+1995,Warin+etal+1996}. However, this method of estimation is influenced by several factors, including the local thermal equilibrium assumption, uncertainties in excitation temperature, and the N(H$_2$)/N(C$^{18}$O) ratio. The uncertainties of H$_2$ column densities in our estimates are generally confined to within an order of magnitude (Paper II). The mass values of the clumps are estimated using the following method: $M=\pi R^2 \mu m_{\text{H}} N(\text{H}_2)$, where $\mu$ is the mean molecular weight of the interstellar medium (i.e., 2.8), and $m{_\text{H}}$ is the mass of a hydrogen atom. Due to uncertainties in distance and radius, the error in clump mass may be relatively large. Therefore, we provide only a rough estimate on the order of magnitude.

The gas infall velocities and mass infall rates for these sources are also provided in Table \ref{Tab:para}, respectively. These values are derived from Papers III and IV, where they are estimated by fitting the line profiles of HCO$^+$ and H$^{13}$CO$^+$. We also listed the HCO$^+$ line profile (the same as these two papers, where B denotes a double-peaked blue profile, P-S denotes a peak-shoulder profile, S denotes a single-peaked profile with the peak skewed to the blue, and N denotes a symmetric line profile or multi-peaked profile) and the number of identified molecular species within the clump area in Table \ref{Tab:para}.

\subsection{Rotation Temperature and Column Density} \label{subsec:para}

We use the LTE radiative transfer model to fit the emission of molecules that were detected in multiple transitions or have a hyperfine structure, in order to estimate physical parameters such as rotation temperatures and column densities. Currently, some studies have shown that using different estimation methods (such as the line intensity equation, rotational diagram, etc.) may lead to differences in the derived rotation temperatures and column densities for certain molecular species. However, for most molecules, the differences in excitation temperatures are generally within 25\%, and column densities within 20\% \citep{Marchand+etal+2024}. In this work, we have adopted the XCLASS program to estimate the molecular parameters. For sources with a clumpy structure, the data used for fitting are the averaged spectra of the area within the clump radius. However, for sources without clear clumpy structures, we have only fit the average spectra of the entire observation area. Firstly, it is essential to identify the factors responsible for the multipeaked line components in optically thick lines. The central radial velocity of the clump is determined using the optically thin line, such as H$^{13}$CO$^+$. For optically thick lines, if multipeaked components exist at corresponding central radial velocities, the line profiles may be caused by self-absorption. Considering that our sources are confirmed infall sources, the optically thick lines generally used to trace gas infall motions, such as HNC, HCO$^+$, and HCN, show complex spectral line profiles that are not suitable for this analysis method. For the species with relatively low abundance in the clump and generally optically thin, we can use the XCLASS program through LTE radiation transfer calculations to generate synthetic spectra that match the observation data.

We focus on molecular lines with SNR $\textgreater$ 3 and presenting multiple transitions within the observed frequency range for fitting, such as c-C$_3$H$_2$, CCH, HNCO, and HC$_3$N. To analyze these spectral lines using XCLASS, the following parameters were required: the size of the clump, its rotation temperature, column density, the FWHM of the spectral lines, and the velocity offset relative to the clump's systemic velocity. The clump size is estimated from section \ref{subsec:size}. The velocity offset and line width are estimated by Gaussian fitting of the lines. We set the initial value of the rotation temperature to 20 K and the molecular column density to approximately 1$\times$10$^{13}$ cm$^{-2}$. To optimize these parameters, we used various algorithms available in MAGIX \citep[Modeling and Analysis Generic Interface for eXternal numerical codes][]{Moller+etal+2013}, including ``genetic", ``Levenberg-Marquardt" and ``errorestim-ins (Interval-Nested-Sampling)" algorithms. This iterative process of parameter adjustment was continued until the spectral line profiles generated by the model match with those observed.

\begin{figure*}
\includegraphics[width=1\textwidth]{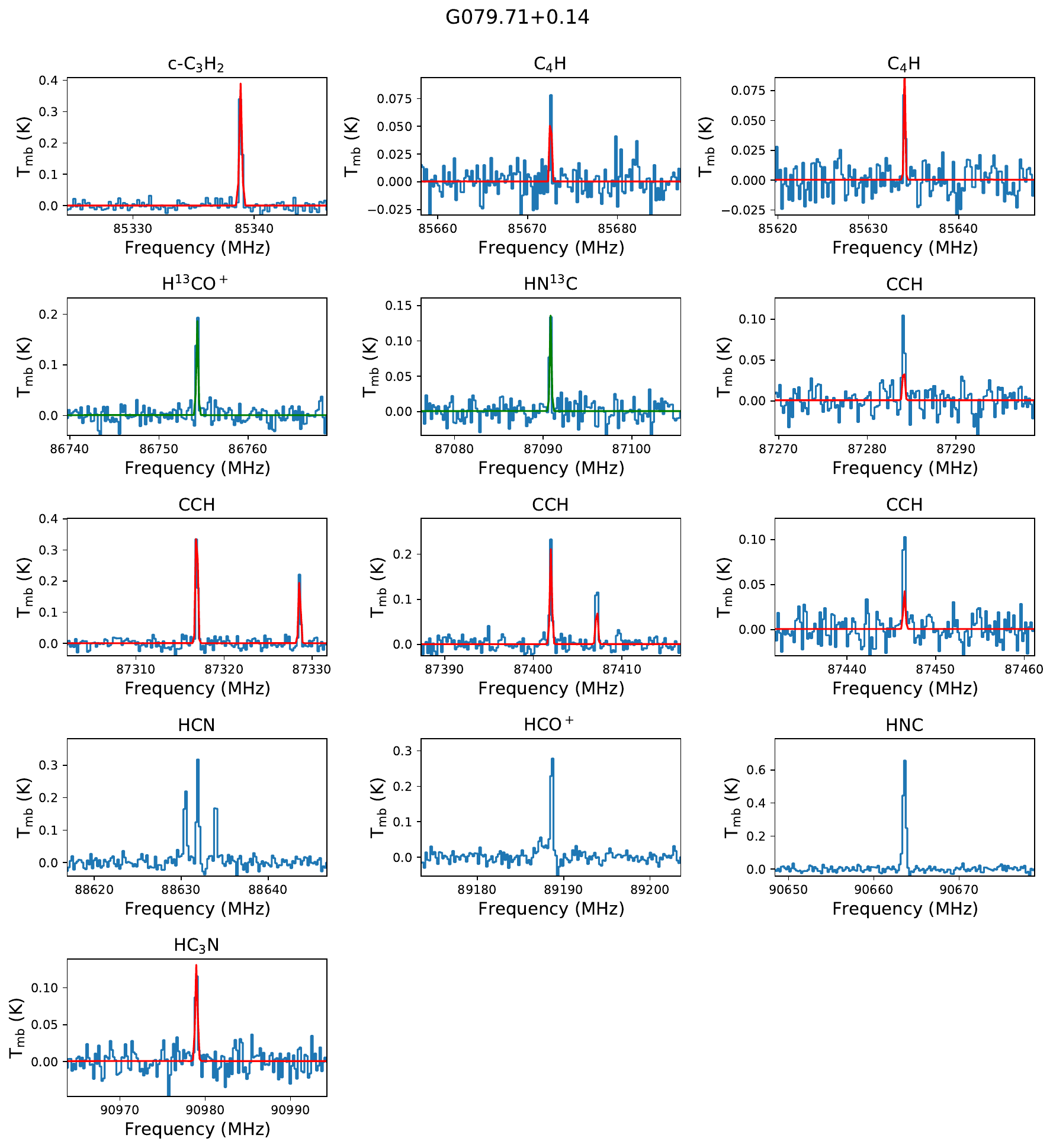}
\caption{Observed spectra (blue) of G079.71+0.14 and the synthesized spectra generated by XCLASS (red), for species observed in this paper. For molecules that exhibit only a single transition line within the observed frequency ranges, and where this line is optically thin, Gaussian fitting is used (green).
\label{fig:mdl}}
\end{figure*}

Figure \ref{fig:mdl} presents the best-fit results of the spectral lines for one of the sources G079.71+0.14 (the complete figure set of all sources are provided online \footnote{\url{https://github.com/yangy4068/2024/blob/Figure-4}}), where the red lines represent the synthesized spectra generated by XCLASS. The best-fit results of rotation temperatures and column densities obtained by this method are listed in Tables \ref{Tab:trot} and \ref{Tab:density}, and the uncertainties are provided by the errorestim-ins algorithm. Due to the limited frequency resolution of our observations, the hyperfine structures of some molecular species could not be well distinguished, and were therefore not applicable to this fitting method. On the other hand, some species show only a single transition line within the observed frequency range. For these molecular species, the same method cannot effectively constrain their parameters.

\begin{figure}
\centering
    \includegraphics[width=0.5\textwidth]{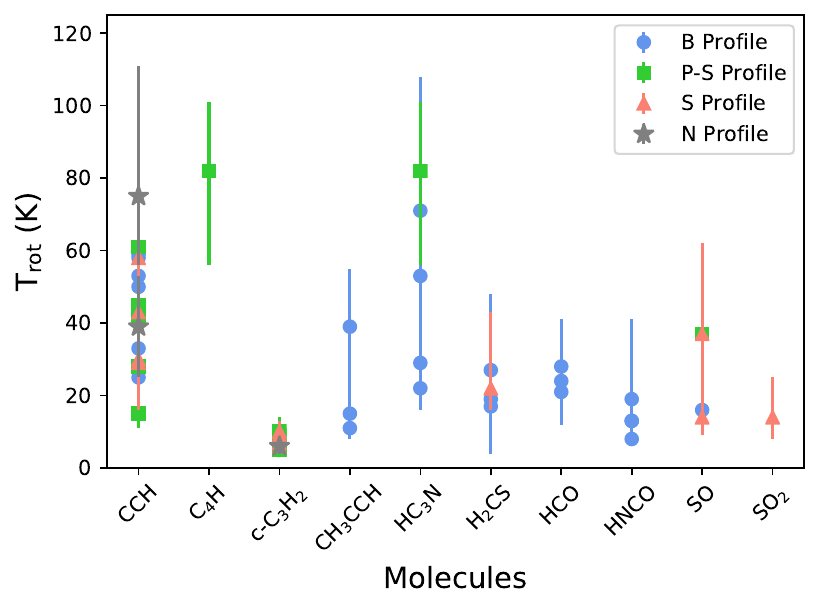}
\caption{The distribution of molecular rotation temperature for the clumps in this paper, with blue dots indicating sources showing a blue profile in the HCO$^+$ lines, green squares indicating sources with peak-shoulder profiles, and red triangles indicating sources with a single-peaked profiles with the peak skewed to the blue. The gray pentagrams indicating a symmetric line profile or a multi-peaked profile.
\label{fig:Trot}}
\end{figure}

The distribution of rotation temperatures for different species is shown in Figure \ref{fig:Trot}, with sources distinguished by their infall line profiles. The temperatures of most molecular species are distributed in the range of tens of Kelvin, while the rotation temperatures of c-C$_3$H$_2$ are mostly below 10 K. For a given species, there is no apparent difference in its temperature distribution, regardless of the infall line profile of the source. 

\begin{table*}
\caption{Molecular rotation temperature of sources}\label{Tab:trot}
\setlength{\tabcolsep}{0.5mm}{
\begin{tabular}{c|ccccccccccccccccccccccc}

  \hline\noalign{\smallskip}
 & G012.79 & G012.87 & G012.87 & G012.96 & G014.25 & G017.09 & G017.09 & G025.82 & G028.97 & G029.06 & G030.17 & G031.41 \\
 & -0.20 & -0.22 A & -0.22 B & -0.23 A & -0.17 & +0.82 A & +0.82 B & -0.18 A & +3.35 & +4.58 & +3.68 & +5.25 \\
 & (K) & (K) & (K) & (K) & (K) & (K) & (K) & (K) & (K) & (K) & (K) & (K) \\
  \hline\noalign{\smallskip}
c-C$_3$H$_2$ & 5$^{+2}_{-2}$ $^1$ & 7$^{+3}_{-1}$ & 7$^{+3}_{-2}$ & 6$^{+2}_{-2}$ & 6$^{+2}_{-2}$ & 8$^{+3}_{-2}$ & 6$^{+2}_{-2}$ &  ...& 10$^{+3}_{-2}$ & 9$^{+3}_{-2}$ &  ...& 10$^{+4}_{-2}$ \\
CH$_3$CCH & 39$^{+16}_{-10}$ & 11$^{+24}_{-3}$ & 15$^{+11}_{-4}$ &  ...&  ...&  ...&  ...&  ...&  ...&  ...&  ...&  ...\\
C$_4$H &  ...&  ...&  ...&  ...&  ...&  ...&  ...&  ...&  ...&  ...&  ...&  ...\\
SO & 16$^{+20}_{-6}$ &  ...&  ...&  ...&  ...&  ...&  ...&  ...& 14$^{+21}_{-5}$ &  ...& 37$^{+25}_{-15}$ &  ...\\
HCO & 24$^{+17}_{-9}$ & 21$^{+15}_{-9}$ & 28$^{+10}_{-10}$ &  ...&  ...&  ...&  ...&  ...&  ...&  ...&  ...&  ...\\
CCH & 33$^{+4}_{-5}$ & 50$^{+15}_{-7}$ & 53$^{+11}_{-14}$ & 59$^{+25}_{-32}$ & 58$^{+5}_{-25}$ & 25$^{+16}_{-8}$ & 43$^{+10}_{-21}$ &  ...& 43$^{+12}_{-16}$ &  ...&  ...& 28$^{+17}_{-9}$ \\
HNCO & 8$^{+5}_{-1}$ & 19$^{+22}_{-6}$ & 13$^{+22}_{-5}$ & 13$^{+23}_{-6}$ &  ...&  ...&  ...&  ...&  ...&  ...&  ...&  ...\\
HC$_3$N & 22$^{+2}_{-4}$ & 53$^{+6}_{-19}$ & 29$^{+5}_{-13}$ &  ...&  ...&  ...&  ...& 71$^{+37}_{-26}$ &  ...&  ...&  ...&  ...\\
H$_2$CS & 17$^{+12}_{-13}$ & 27$^{+21}_{-8}$ & 19$^{+22}_{-7}$ &  ...&  ...&  ...&  ...&  ...& 22$^{+21}_{-6}$ &  ...&  ...&  ...\\
SO$_2$ &  ...&  ...&  ...&  ...&  ...&  ...&  ...&  ...& 14$^{+11}_{-6}$ &  ...&  ...&  ...\\
  \hline\noalign{\smallskip}
 & G036.02 & G037.05 & G049.07 & G049.07 & G079.71 & G121.34 & G126.53 & G133.42 & G133.42 & G143.04 & G154.05 \\
 & -1.36 A & -0.03 & -0.33 A & -0.33 B & +0.14 & +3.42 & -1.17 & +0.00 A & +0.00 C & +1.74 & +5.07 \\
 & (K) & (K) & (K) & (K) & (K) & (K) & (K) & (K) & (K) & (K) & (K) &  \\
  \hline\noalign{\smallskip}
c-C$_3$H$_2$ & 5$^{+3}_{-2}$ & 5$^{+2}_{-2}$ & 9$^{+3}_{-2}$ & 6$^{+2}_{-2}$ & 5$^{+3}_{-2}$ & 6$^{+3}_{-1}$ & 10$^{+3}_{-2}$ &  ...&  ...& 7$^{+3}_{-2}$ &  ...&  \\
CH$_3$CCH &  ...&  ...&  ...&  ...&  ...&  ...&  ...&  ...&  ...&  ...&  ...&  \\
C$_4$H &  ...&  ...&  ...&  ...& 82$^{+19}_{-26}$ &  ...&  ...&  ...&  ...&  ...&  ...&  \\
SO &  ...&  ...&  ...&  ...&  ...&  ...&  ...&  ...&  ...&  ...& 20$^{+20}_{-6}$ &  \\
HCO &  ...&  ...&  ...&  ...&  ...&  ...&  ...&  ...&  ...&  ...&  ...&  \\
CCH & 41$^{+37}_{-18}$ & 15$^{+8}_{-4}$ & 45$^{+17}_{-9}$ & 75$^{+36}_{-19}$ & 61$^{+5}_{-42}$ & 29$^{+16}_{-10}$ & 30$^{+30}_{-14}$ & 43$^{+16}_{-13}$ & 39$^{+14}_{-14}$ & 58$^{+10}_{-20}$ &  ...&  \\
HNCO &  ...&  ...&  ...&  ...&  ...&  ...&  ...&  ...&  ...&  ...&  ...&  \\
HC$_3$N &  ...&  ...&  ...&  ...& 40$^{+20}_{-10}$ &  ...&  ...&  ...&  ...&  ...&  ...&  \\
H$_2$CS &  ...&  ...&  ...&  ...&  ...&  ...&  ...&  ...&  ...&  ...&  ...&  \\
SO$_2$ &  ...&  ...&  ...&  ...&  ...&  ...&  ...&  ...&  ...&  ...&  ...&  \\
  \hline\noalign{\smallskip}
\end{tabular}}
\footnotesize{$^1$ The best-fit results of rotation temperatures obtained by XCLASS, and the uncertainties are provided by the errorestim-ins algorithm.}
\end{table*}

\begin{table*}
\caption{Molecular column densities and abundances of sources}\label{Tab:density}
\setlength{\tabcolsep}{0.05mm}{
\hspace*{-1.5cm} 
\begin{tabular}{c|cccccccccccc}

  \hline\noalign{\smallskip}
 & \multicolumn{2}{c}{G012.79-0.20} & \multicolumn{2}{c}{G012.87-0.22 A} & \multicolumn{2}{c}{G012.87-0.22 B} & \multicolumn{2}{c}{G012.96-0.23 A} & \multicolumn{2}{c}{G014.25-0.17} & \multicolumn{2}{c}{G017.09+0.82 A} \\
 & N(mole) & X(mole) & N(mole) & X(mole) & N(mole) & X(mole) & N(mole) & X(mole) & N(mole) & X(mole) & N(mole) & X(mole) \\
 & (10$^{12}$ cm$^{-2}$) & (10$^{-10}$) & (10$^{12}$ cm$^{-2}$) & (10$^{-10}$) & (10$^{12}$ cm$^{-2}$) & (10$^{-10}$) & (10$^{12}$ cm$^{-2}$) & (10$^{-10}$) & (10$^{12}$ cm$^{-2}$) & (10$^{-10}$) & (10$^{12}$ cm$^{-2}$) & (10$^{-10}$) \\
  \hline\noalign{\smallskip}
c-C$_3$H$_2$ & 36$^{+13}_{-12}$ $^1$ & 2.7$^{+1.1}_{-1.0}$  $^2$ & 9.5$^{+0.1}_{-0.5}$ & 0.54$^{+0.11}_{-0.11}$ & 11$^{+2}_{-5}$ & 1.1$^{+0.3}_{-0.5}$  & 7.7$^{+0.2}_{-0.5}$ & 0.66$^{+0.13}_{-0.14}$ & 6.3$^{+0.6}_{-0.3}$ & 0.89$^{+0.20}_{-0.18}$ & 5.0$^{+1.0}_{-0.2}$ & 1.3$^{+0.4}_{-0.3}$ \\
CH$_3$CCH & 1000$^{+40}_{-80}$ & 74$^{+15}_{-16}$ & 60$^{+9}_{-3}$ & 3.4$^{+0.9}_{-0.7}$ & 91$^{+1}_{-2}$ & 8.9$^{+1.8}_{-1.8}$ &  ...&  ... &  ...&  ... &  ...&  ...\\
C$_4$H &  ...&  ... &  ...&  ... &  ...&  ... &  ...&  ... &  ...&  ... &  ...&  ...\\
SO & 190$^{+20}_{-46}$ & 14 $^{+3}_{-4}$&  ...&  ... &  ...&  ... &  ...&  ... &  ...&  ... &  ...&  ...\\
HCO & 55$^{+9}_{+7}$ & 4.1$^{+1.1}_{-1.0}$ & 15$^{+8}_{-3}$ & 0.85$^{+0.48}_{-0.23}$ & 31$^{+2}_{-5}$ & 3.1$^{+0.6}_{-0.8}$ &  ...&  ... &  ...&  ... &  ...&  ...\\
CCH & 3100$^{+60}_{-110}$ & 230$^{+47}_{-47}$ & 1000$^{+50}_{-250}$ & 59$^{+12}_{-19}$ & 680$^{+11}_{-34}$ & 67$^{+13}_{-14}$ & 360$^{+5}_{-80}$ & 30$^{+6}_{-9}$ & 220$^{+29}_{-39}$ & 31$^{+8}_{-8}$ & 260$^{+37}_{-68}$ & 66$^{+16}_{-22}$ \\
HNCO & 21$^{+2}_{-5}$ & 1.6$^{+0.4}_{-0.5}$ & 5.6$^{+0.2}_{-0.4}$ & 0.32$^{+0.07}_{-0.07}$ & 7.0$^{+0.1}_{-0.4}$ & 0.69$^{+0.14}_{-0.14}$ & 3.3$^{+1.2}_{-0.2}$ & 0.28$^{+0.12}_{-0.06}$ &  ...&  ... &  ...&  ...\\
HC$_3$N & 28$^{+1}_{-1}$ & 2.0$^{+0.4}_{-0.4}$ & 3.7$^{+0.8}_{-0.3}$ & 0.21$^{+0.06}_{-0.05}$ & 1.4$^{+0.2}_{-0.3}$ & 0.13$^{+0.03}_{-0.04}$ &  ...&  ... &  ...&  ... &  ...&  ...\\
H$_2$CS & 130$^{+7}_{-21}$ & 9.4$^{+2.0}_{-2.4}$ & 27$^{+2}_{-5}$ & 1.6$^{+0.3}_{-0.4}$ & 10$^{+9}_{-2}$ & 1.0$^{+0.9}_{-0.3}$ &  ...&  ... &  ...&  ... &  ...&  ...\\
SO$_2$ &  ...&  ... &  ...&  ... &  ...&  ... &  ...&  ... &  ...&  ... &  ...&  ...\\
  \hline\noalign{\smallskip}
 & \multicolumn{2}{c}{G017.09+0.82 B} & \multicolumn{2}{c}{G025.82-0.18 A} & \multicolumn{2}{c}{G028.97+3.35} & \multicolumn{2}{c}{G029.06+4.58} & \multicolumn{2}{c}{G030.17+3.68} & \multicolumn{2}{c}{G031.41+5.25} \\
 & N(mole) & X(mole) & N(mole) & X(mole) & N(mole) & X(mole) & N(mole) & X(mole) & N(mole) & X(mole) & N(mole) & X(mole) \\
 & (10$^{12}$ cm$^{-2}$) & (10$^{-10}$) & (10$^{12}$ cm$^{-2}$) & (10$^{-10}$) & (10$^{12}$ cm$^{-2}$) & (10$^{-10}$) & (10$^{12}$ cm$^{-2}$) & (10$^{-10}$) & (10$^{12}$ cm$^{-2}$) & (10$^{-10}$) & (10$^{12}$ cm$^{-2}$) & (10$^{-10}$) \\
  \hline\noalign{\smallskip}
c-C$_3$H$_2$ & 5.0$^{+1.0}_{-0.2}$ & 1.3$^{+0.4}_{-0.3}$ &  ...&  ... & 2.0$^{+0.3}_{-0.4}$ & 0.81$^{+0.19}_{-0.23}$ & 2.5$^{+0.2}_{-0.5}$ & 6.2$^{+1.3}_{-1.7}$ &  ...&  ... & 0.49$^{+0.05}_{-0.02}$ & 0.49$^{+0.11}_{-0.11}$ \\
CH$_3$CCH &  ...&  ... &  ...&  ... &  ...&  ... &  ...&  ... &  ...&  ... &  ...&  ...\\
C$_4$H &  ...&  ... &  ...&  ... &  ...&  ... &  ...&  ... &  ...&  ... &  ...&  ...\\
SO &  ...&  ... &  ...&  ... & 9.6$^{+0.3}_{-0.4}$ & 3.8$^{+0.8}_{-0.8}$ &  ...&  ... & 23$^{+2}_{-4}$ & 36$^{+8}_{-10}$ &  ...&  ...\\
HCO &  ...&  ... &  ...&  ... &  ...&  ... &  ...&  ... &  ...&  ... &  ...& \\
CCH & 710$^{+10}_{-35}$ & 180$^{+37}_{-38}$ &  ...&  ... & 120$^{+80}_{-46}$ & 47$^{+33}_{-20}$ &  ...&  ... &  ...&  ... & 25$^{+2}_{-4}$ & 25$^{+6}_{-7}$ \\
HNCO &  ...&  ... &  ...&  ... &  ...&  ... &  ...&  ... &  ...&  ... &  ...&  ...\\
HC$_3$N &  ...&  ... & 8.2$^{+1.0}_{-0.3}$ & 1.5$^{+0.3}_{-0.3}$ &  ...&  ... &  ...&  ... &  ...&  ... &  ...&  ...\\
H$_2$CS &  ...&  ... &  ...&  ... & 2.7$^{+0.2}_{+0.5}$ & 1.1$^{+0.2}_{-0.3}$ &  ...&  ... &  ...&  ... &  ...&  ...\\
SO$_2$ &  ...&  ... &  ...&  ... & 4.3$^{+0.1}_{-0.6}$ & 1.7$^{+0.4}_{-0.4}$ &  ...&  ... &  ...&  ... &  ...&  ...\\
  \hline\noalign{\smallskip}
 & \multicolumn{2}{c}{G036.02-1.36 A} & \multicolumn{2}{c}{G037.05-0.03} & \multicolumn{2}{c}{G049.07-0.33 A} & \multicolumn{2}{c}{G049.07-0.33 B} & \multicolumn{2}{c}{G079.71+0.14} & \multicolumn{2}{c}{G121.34+3.42} \\
 & N(mole) & X(mole) & N(mole) & X(mole) & N(mole) & X(mole) & N(mole) & X(mole) & N(mole) & X(mole) & N(mole) & X(mole) \\
 & (10$^{12}$ cm$^{-2}$) & (10$^{-10}$) & (10$^{12}$ cm$^{-2}$) & (10$^{-10}$) & (10$^{12}$ cm$^{-2}$) & (10$^{-10}$) & (10$^{12}$ cm$^{-2}$) & (10$^{-10}$) & (10$^{12}$ cm$^{-2}$) & (10$^{-10}$) & (10$^{12}$ cm$^{-2}$) & (10$^{-10}$) \\
  \hline\noalign{\smallskip}
c-C$_3$H$_2$ & 5.5$^{+0.4}_{-0.3}$ & 6.5$^{+1.4}_{-1.3}$ & 7.4$^{+0.8}_{-0.2}$ & 2.3$^{+0.5}_{-0.5}$ & 9.3$^{+1.0}_{-0.2}$ & 3.1$^{+0.7}_{-0.6}$ & 18$^{+2}_{-4}$ & 6.0$^{+1.4}_{-1.9}$ & 3.7$^{+0.5}_{-0.3}$ & 7.4$^{+1.7}_{-1.6}$ & 0.95$^{+0.04}_{-0.03}$ & 0.95$^{+0.19}_{-0.19}$ \\
CH$_3$CCH &  ...&  ...&  ...&  ... &  ...&  ... &  ...&  ... &  ...&  ... &  ...&  ...\\
C$_4$H &  ...&  ...&  ...&  ... &  ...&  ... &  ...&  ... & 31$^{+2}_{-5}$ & 61$^{+38}_{-16}$ &  ...&  ...\\
SO &  ...&  ...&  ...&  ... &  ...&  ... &  ...&  ... &  ...&  ... &  ...&  ...\\
HCO &  ...&  ...&  ...&  ... &  ...&  ... &  ...&  ... &  ...&  ... &  ...&  ...\\
CCH & 55$^{+2}_{-4}$ & 66$^{+13}_{-14}$ & 150$^{+19}_{-47}$ & 47$^{+11}_{-17}$ & 990$^{+8}_{-26}$ & 330$^{+66}_{-66}$ & 1300$^{+190}_{-480}$ & 430$^{+110}_{-180}$ & 150$^{+14}_{-50}$ & 300$^{+66}_{-120}$ & 18$^{+3}_{-4}$ & 18$^{+5}_{-5}$ \\
HNCO &  ...&  ...&  ...&  ... &  ...&  ... &  ...&  ... &  ...&  ... &  ...&  ...\\
HC$_3$N &  ...&  ...&  ...&  ... &  ...&  ... &  ...&  ... & 0.47$^{+0.10}_{-0.02}$ & 0.94$^{+0.27}_{-0.19}$ &  ...&  ...\\
H$_2$CS &  ...&  ...&  ...&  ... &  ...&  ... &  ...&  ... &  ...&  ... &  ...&  ...\\
SO$_2$ &  ...&  ...&  ...&  ... &  ...&  ... &  ...&  ... &  ...&  ... &  ...&  ...\\
  \hline\noalign{\smallskip}
\end{tabular}}
\end{table*}

\begin{table*}
\renewcommand{\thetable}{6}
\caption{Continued.}\label{Tab:density}
\setlength{\tabcolsep}{0.05mm}{
\hspace*{-1.5cm} 
\begin{tabular}{c|cccccccccccc}

  \hline\noalign{\smallskip}
 & \multicolumn{2}{c}{G126.53-1.17} & \multicolumn{2}{c}{G133.42+0.00 A} & \multicolumn{2}{c}{G133.42+0.00 C} & \multicolumn{2}{c}{G143.04+1.74} & \multicolumn{2}{c}{G154.05+5.07}  & \phantom{GXXX.XX+X.XX} & \phantom{Column 0} \\
 & N(mole) & X(mole) & N(mole) & X(mole) & N(mole) & X(mole) & N(mole) & X(mole) & N(mole) & X(mole) & \null & \null \\
 & (10$^{12}$ cm$^{-2}$) & (10$^{-10}$) & (10$^{12}$ cm$^{-2}$) & (10$^{-10}$) & (10$^{12}$ cm$^{-2}$) & (10$^{-10}$) & (10$^{12}$ cm$^{-2}$) & (10$^{-10}$) & (10$^{12}$ cm$^{-2}$) & (10$^{-10}$) & \null & \null \\
  \hline\noalign{\smallskip}
c-C$_3$H$_2$ & 3.2$^{+0.1}_{-0.6}$ & 3.2$^{+0.7}_{-0.9}$ &  ...&  ... &  ...&  ...& 3.6$^{+0.1}_{-0.6}$ & 3.6$^{+0.7}_{-0.9}$ &  ...&  ...& \null & \null \\
CH$_3$CCH &  ...&  ... &  ...&  ... &  ...&  ...&  ...&  ... &  ...&  ...& \null & \null \\
C$_4$H &  ...&  ... &  ...&  ... &  ...&  ...&  ...&  ... &  ...&  ...& \null & \null \\
SO &  ...&  ... &  ...&  ... &  ...&  ...&  ...&  ... & 9.4$^{+0.3}_{-0.8}$ & 30$^{+6}_{-7}$ & \null & \null \\
HCO &  ...&  ... &  ...&  ... &  ...&  ...&  ...&  ... &  ...&  ...& \null & \null \\
CCH & 22$^{+2}_{-8}$ & 22$^{+5}_{-9}$  & 69$^{+11}_{-2}$ & 140$^{+29}_{-19}$ & 78$^{+11}_{-18}$ & 160$^{+40}_{-50}$ & 100$^{+58}_{-22}$ & 100$^{+62}_{-30}$  &  ...&  ...& \null & \null \\
HNCO &  ...&  ... &  ...&  ... &  ...&  ...&  ...&  ... &  ...&  ...& \null & \null \\
HC$_3$N &  ...&  ... &  ...&  ... &  ...&  ...&  ...&  ... &  ...&  ...& \null & \null \\
H$_2$CS &  ...&  ... &  ...&  ... &  ...&  ...&  ...&  ... &  ...&  ...& \null & \null \\
SO$_2$ &  ...&  ... &  ...&  ... &  ...&  ...&  ...&  ... &  ...&  ...& \null & \null \\
  \hline\noalign{\smallskip}
\end{tabular}}
\footnotesize{$^1$ The best-fit results of column densities obtained by XCLASS, with uncertainties provided by the errorestim-ins algorithm. $^2$ The corresponding molecular abundances, with uncertainties derived from error propagation.}
\end{table*}

\subsection{Abundance} \label{subsec:abund}

\subsubsection{Molecular Abundances of the Clumps} \label{subsubsec:abund1}

\begin{figure}
\centering
    \includegraphics[width=0.49\textwidth]{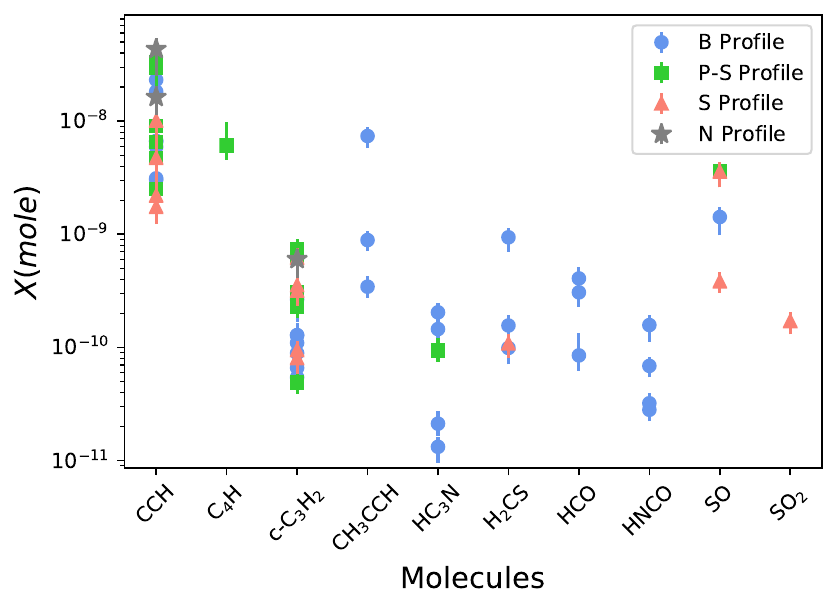}
    \includegraphics[width=0.49\textwidth]{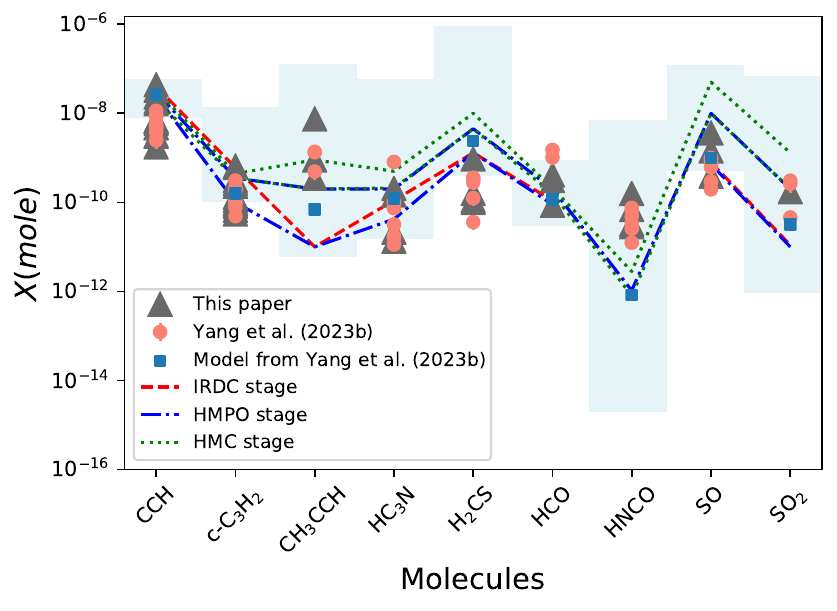}
\caption{Left panel: The distribution of molecular abundances for the clumps in this paper. Different markers represent the molecular abundances from clumps with different infall line profiles (the same as Figure \ref{fig:Trot}). Right panel: The distribution of molecular abundances obtain in this paper (grey triangles) and in Paper V (red dots). The blue squares represent the simulated average molecular abundances for the HMPO stage from the chemical model. The red dashed line, blue dash-dot line and green dotted line represent the maximum and minimum average abundances of various molecular species at the IRDC, HMPO and hot molecular core (HMC) stages simulated by the model, respectively. The light blue shaded regions represent the range of molecular abundances at different locations within the clump in HMPO stage.
\label{fig:abundances}}
\end{figure}

Based on the estimated H$_2$ column densities provided above, the abundance of each observed molecular species in these sources is calculated using the formula $X(\text{mole}) = N(\text{mole})/N(\text{H}_2)$, where the column density of the molecular species is the best-fit column density value obtained from MAGIX (see Table \ref{Tab:density}). Due to the estimation method of H$_2$ column density, $N(\text{H}_2)$ values may be subject to errors by a factor of several. Therefore, the actual errors in molecular abundances may exceed the error ranges we have provided here. Through our multiple tests, such errors will not lead to differences in the order of magnitude of the abundances in most cases. The left panel of Figure \ref{fig:abundances} shows the distribution of all molecular abundances. Here, different markers represent the molecular abundances from clumps with different infall line profiles. The distribution of molecular abundances does not seem to depend on the type of infall profile. However, the H$_2$ column densities estimated from previous studies indicate that these sources with typical blue profiles tend to have slightly higher H$_2$ column densities than those with other infall line profiles. 

\subsubsection{Comparison with nine star-forming regions and the model} \label{subsubsec:Compar}

Paper V reported the molecular distributions in nine high-mass star-forming regions with typical blue profile. In addition, we provided the molecular abundances of these nine sources and compared them with the values simulated by a chemical model, which adopted physical conditions of high-mass star-forming regions \citep{Gerner+etal+2014, Gerner+etal+2015} to fit the observed molecular abundances. The two-phase gas-grain model was adopted in this model \citep{Hasegawa+etal+1992}, and the reaction network was adopted from \citet{Wang+etal+2021}. The comparison results indicated that these nine sources are in the early HMPO stage, with the inner temperature around several 10 K and approximately 1.7 to 2.0 $\times$ 10$^4$ years. 

We compared the molecular abundances of these nine sources with those reported in this study. Firstly, for G029.06+4.58, G031.41+5.25, G079.71+0.14, G126.53-1.17, G133.42+0.00 clump A, G143.04+1.74, and G154.05+5.07, the estimated masses of these sources are relatively low, suggesting that they may belong to intermediate or low-mass star-forming regions. Therefore, they are not included in the discussion. For the remaining sources, we present a comparison of their molecular abundances with those of previously reported nine sources in the right panel of Figure \ref{fig:abundances}. As shown in this figure, the grey triangles represent sources from this study, marking only the molecules that were detected and for which abundances could be derived in both studies. The red dots represent the nine infall sources from Paper V. The blue squares represent the simulated average abundances for the HMPO stage from the chemical model. The abundance distribution of the other molecules in the sources in this paper is similar to that in the sources in paper V, except that the HCO abundance is slightly lower. Additionally, we use lines to indicate the maximum and minimum abundances of various molecular species at different evolutionary stages simulated by the model. Since the simulation of the infrared dark cloud (IRDC) stage did not provide the minimum values, only the upper limits of the simulated values were plotted. The different line styles in the figure represent the abundances of different evolutionary stages. 

We mainly compared the IRDC and HMPO stages since these sources are undergoing gas infall. For G012.79-0.20, G012.87-0.22 clump A and clump B, and G012.96-0.23 clump A, their molecular abundances of most species are consistent with the abundances of clumps that have evolved to the HMPO stage, as provided by the model, except for CH$_3$CCH, HNCO, and H$_2$CS. For other high-mass infall sources where only a few species, such as CCH and c-C$_3$H$_2$, are detected, we can only speculate that they may be in the earlier IRDC stage, during which many molecular species have not yet been released from the dust. However, we still need molecular data from other bands to further confirm.

As previously mentioned, there are still differences between the simulated and observed values for a few molecular abundances. For the clumps that may be in the HMPO stages, the observed abundances of CH$_3$CCH and HNCO are higher than the simulated results, while the observed abundances of H$_2$CS are lower than the simulated results. One possibility is that the real sources have complex clump structures and physical environments, while the model cannot simulate the complex spatial distribution of molecular species. In the right panel of Figure \ref{fig:abundances}, we have marked the range of molecular abundances at different locations within the clumps provided by the model with light blue shaded regions. It can be seen that although the average abundances, such as H$_2$CS, cannot fit observations well, the abundances in different locations deviate from observations within one order of magnitude. Future observations with higher sensitivity and spatial resolution are needed to confirm whether these species are distributed in different regions. Another possibility is that the coefficients of related reactions may be inaccurate, affecting their abundances, as well as the reaction network may be incomplete for producing CH$_3$CCH and HNCO and destroying H$_2$CS. Future improvements in the chemical network may reduce discrepancies between model and observations.

\subsection{Comparison of CCH and c-C$_3$H$_2$ column densities} \label{subsec:cch}

\begin{figure}
\includegraphics[width=3in]{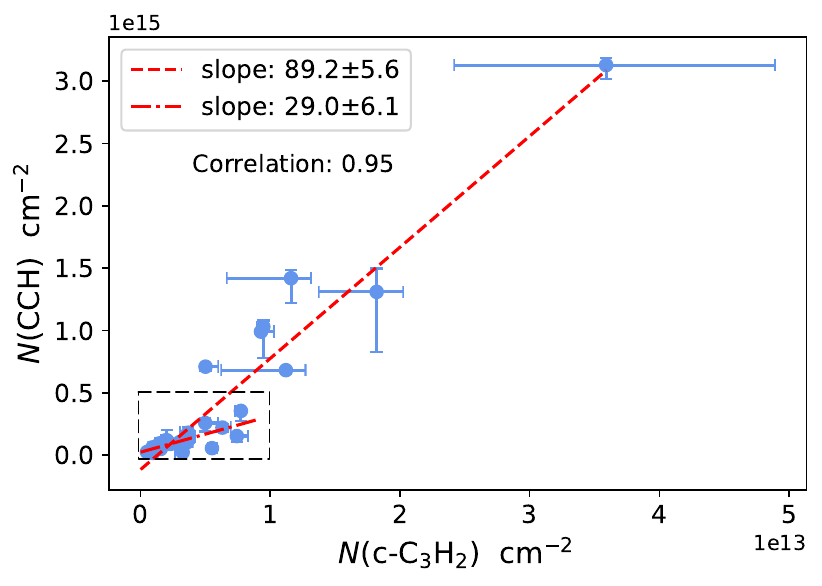}
\includegraphics[width=3in]{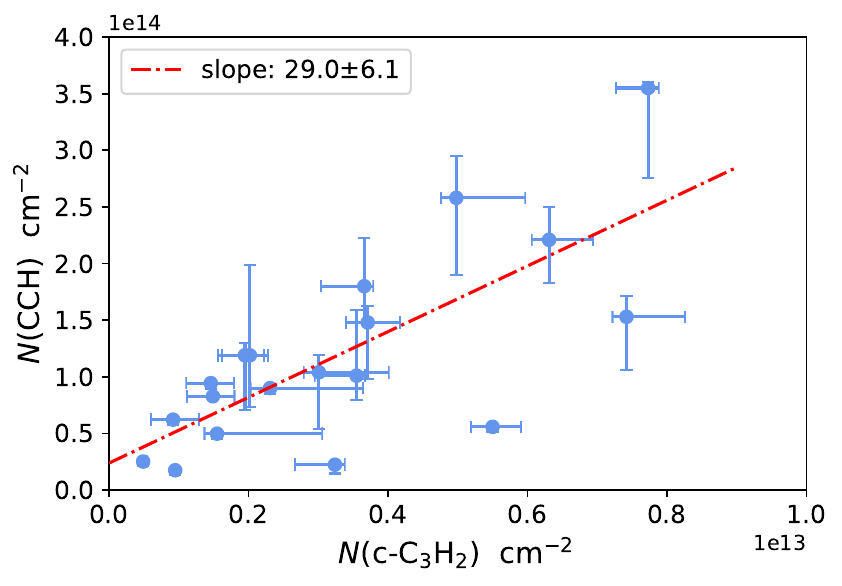}
\caption{
Left panel: Comparison of CCH and c-C$_3$H$_2$ column densities for all sources. The red dashed line represents a slope of approximately 89.2. Right panel: Comparison of CCH and c-C$_3$H$_2$ column densities in the lower density region highlighted by black dashed lines in the left panel. The red dash-dotted line represents a slope of approximately 29.0, which is consistent with the correlation found by \citet{Lucas+Liszt+2000}, \citet{Gerin+Kazmierczak+2011}, \citet{Schmidt+etal+2018}, and \citet{Qiu+etal+2023}.
\label{fig:cch_c3h2}}
\end{figure}

\begin{figure}
\includegraphics[width=3in]{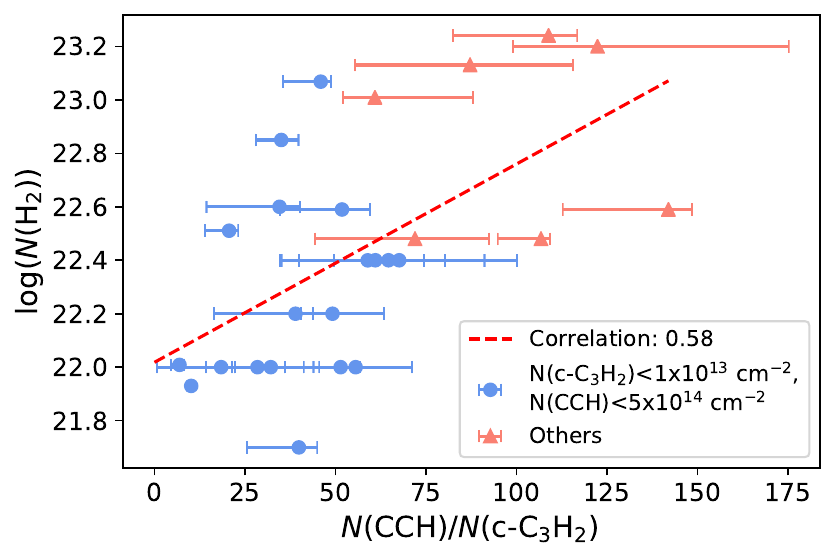}
\includegraphics[width=3in]{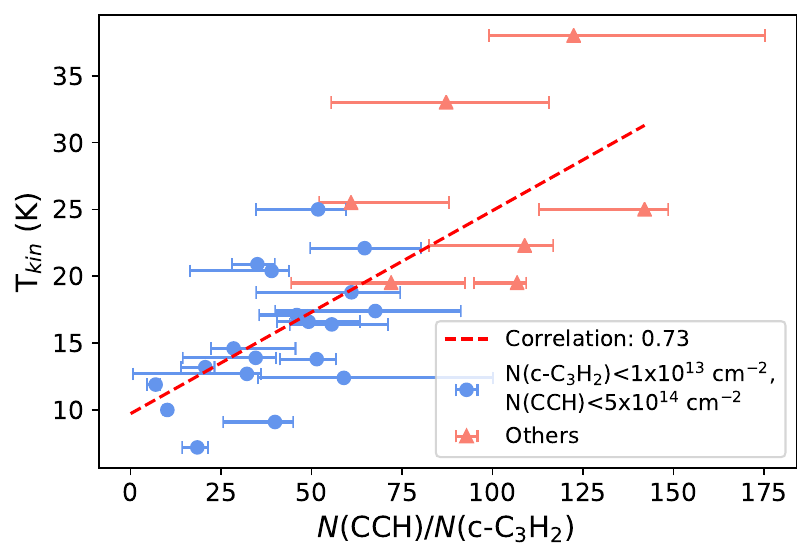}
\caption{Left panel: The correlation between the N(CCH)/N(c-C$_3$H$_2$) ratios and the H$_2$ column densities of sources. Right panel: The correlation between the N(CCH)/N(c-C$_3$H$_2$) ratios and the kinetic temperatures of sources. In both panels, blue dots represent the sources shown in the right panel of Figure \ref{fig:cch_c3h2}, while red triangles represent the remaining sources. The red dashed line indicates the correlation line in each panel.
\label{fig:cch_c3h2_2}}
\end{figure}

Analysis of column densities of CCH and c-C$_3$H$_2$ in the observed sources (including the nine sources reported in Paper V) reveals a linear correlation between them, with a correlation coefficient of 0.95. This is because the formation and destruction routes between these two molecules are associated \citep{Qiu+etal+2023}. This correlation is shown in the left panel of Figure \ref{fig:cch_c3h2}. By conducting a linear fit to the column densities of CCH and c-C$_3$H$_2$, we obtained a ratio of N(CCH)/N(c-C$_3$H$_2$) = 89.2$\pm$5.6. This ratio is higher than those reported in literature: 27.7$\pm$8 for local diffuse clouds \citep{Lucas+Liszt+2000}, 28.2$\pm$1.4 for diffuse and translucent clouds \citep{Gerin+Kazmierczak+2011}, 29$\pm$6 for old planetary nebulae \citep{Schmidt+etal+2018}, 23.2$\pm$1.9 for a possible red nova remnant \citep{Qiu+etal+2023}. However, when focusing on sources with lower column densities (with c-C$_3$H$_2$ column density approximately less than 1$\times$10$^{13}$ cm$^{-2}$ and CCH column density approximately less than 5$\times$10$^{14}$ cm$^{-2}$), the ratio is approximately 29.0$\pm$6.1, as shown in the right panel of Figure \ref{fig:cch_c3h2}. While the overall ratio of N(CCH)/N(c-C$_3$H$_2$) in our sources is significantly higher than the values reported in the literature, the subset of sources with lower CCH and c-C$_3$H$_2$ column densities exhibits a ratio consistent with the values found in diffuse clouds. 

To further explore their physical conditions, we plot the correlation between the N(CCH)/N(c-C$_3$H$_2$) ratios and the physical properties of these sources, as shown in Figure \ref{fig:cch_c3h2_2}. The figure shows that sources with lower N(CCH)/N(c-C$_3$H$_2$) ratios have slightly lower average H$_2$ column densities compared to other sources. Their kinetic temperatures are also relatively lower than those of other sources, mostly below 25 K. Sources with higher N(CCH)/N(c-C$_3$H$_2$) ratios tend to have higher H$_2$ column densities and kinetic temperatures. The correlation coefficient between the N(CCH)/N(c-C$_3$H$_2$) ratios and H$_2$ column densities is 0.58, and that between the ratios and kinetic temperatures is 0.73. Additionally, some studies have suggested that N(CCH)/N(c-C$_3$H$_2$) may be linearly related to $\chi/n_{\mathrm{H}}$ ($\chi$ denotes the radiation field in Draine units), and it decreases when the UV field strength decreases \citep{Cuadrado+etal+2015}. The higher ratio we observed may imply that these sources are exposed to more intense UV radiation fields. However, this still requires more observational data to support.

\section{Conclusions}  \label{sec:summary}

Using the FTS wide-sideband mode of the IRAM 30 m telescope, we observed 20 sources with gas infall motion at frequency ranges of 83.7 -- 91.5 GHz and 99.4 -- 107.2 GHz. In this study, we identify multiple molecular species in these sources, and use these data to analyse the physical and chemical properties of these infall sources. The conclusions are as follows:

(i) Using XCLASS, 22 molecular species and their isotopologue transition lines, along with one hydrogen radio recombination line, have been identified in 20 infall sources. Among them, 15 sources have lines of H$^{13}$CO$^+$, HCO$^+$, HCN, HNC, c-C$_3$H$_2$, and CCH detected. The source with the highest number of molecular lines is G012.79-0.20, while G110.40+1.67 has the fewest molecular emission lines.

(ii) In most cases, integrated intensity maps reveal spatially correlated compact regions for various molecular species. 24 clumps have been identified in these observed sources, and four sources show no discernible clumpy structures. Additionally, the physical properties of these clumps are estimated.

(iii) We use XCLASS's LTE radiative transfer calculations to estimate the molecular rotation temperatures and column densities, and compare these molecular column densities with the H$_2$ column density to obtain the abundance of each molecular species in these sources. Sources with different infall profiles, such as blue profile, peak-shoulder profile and blue-skewed single-peaked profile, do not show significant differences in the distribution of the same molecular abundances. 

(iv) By comparing the sources in this study with nine high-mass star-forming regions reported in previous studies and the chemical model, we find that G012.79-0.20, G012.87-0.22 clump A and clump B, and G012.96-0.23 clump A may consistent with the clumps that have evolved to the HMPO stage, as provided by the model. For other high-mass sources where only a few species, such as CCH and c-C$_3$H$_2$, are detected, they may be in the IRDC stage.

(v) The column densities of CCH and c-C$_3$H$_2$ of our sources reveals a linear correlation, with a ratio of N(CCH)/N(c-C$_3$H$_2$) = 89.2$\pm$5.6. The ratio in sources with lower column densities (i.e. 29.0$\pm$6.1) is consistent with values reported for diffuse clouds.


\normalem
\begin{acknowledgements}

We are grateful to the staff of the Institut de Radioastronomie Millim{\'e}trique (IRAM) for their assistance and support during the observations. This work has been supported by the National Key R\&D Program of China (No. 2022YFA1603102), and the National Natural Science Foundation of China (NSFC) Grant Nos. U2031202, 12373030, 11873093.

\end{acknowledgements}




\bibliographystyle{raa}
\bibliography{RAA-2025-0056.R1}

\end{document}